\documentclass[structabstract]{aa}

\usepackage{graphicx}
\usepackage{txfonts, longtable, dcolumn, ctable,verbatim}
\usepackage[T1]{fontenc}
\usepackage{natbib}
\usepackage{bibentry}
\nobibliography*
\bibpunct{(}{)}{;}{a}{}{,}

\usepackage[squaren,textstyle]{SIunits}

\newcommand{\tiu}{\joule\usk\power{\metre}{-2}\power{\second}{-1/2}\power{\kelvin}{-1}}

\usepackage{color}
\usepackage{amstext}

\begin{document}

   \title{Near-Earth asteroid (3200) Phaethon. Characterization of its orbit, spin state, and thermophysical parameters}

   \author{J. Hanu{\v s}
          \inst{1,2*}
           \and
          M.~Delbo'\inst{2}
           \and
          D.~Vokrouhlick\'y\inst{3}
           \and
          P.~Pravec\inst{4}
           \and
          J.P.~Emery\inst{5}
           \and
              V.~Al\'i-Lagoa\inst{2}
           \and           
          B.~Bolin\inst{2}
           \and           
          M.~Devog\`{e}le\inst{6,2}
           \and     
          R.~Dyvig\inst{7}
           \and
          A.~Gal\'ad\inst{8}
           \and
          R.~Jedicke\inst{9}
           \and
          L.~Korno\v s\inst{8}
           \and
          P.~Ku\v snir\'ak\inst{4}
           \and
          J.~Licandro\inst{10,11}
           \and
          V.~Reddy\inst{12}
           \and
          J-P.~Rivet\inst{2}
           \and           
          J.~Vil\' agi\inst{8}
           \and
          B.D.~Warner\inst{13}
}

   \institute{
             Centre National d'\'Etudes Spatiales, 2 place Maurice Quentin, 75039 Paris cedex 01, France\\
             $^*$\email{hanus.home@gmail.com}
         \and
             Laboratoire Lagrange, UMR7293, Universit\' e de la C\^ ote d'Azur, CNRS, Observatoire de la C\^ ote d'Azur, Blvd de l'Observatoire, CS 34229, 06304 Nice cedex 04, France
         \and
         Astronomical Institute, Faculty of Mathematics and Physics, Charles University, V~Hole\v sovi\v ck\' ach 2, 180 00 Prague, Czech Republic
         \and
             Astronomical Institute, Academy of Sciences of the Czech Republic, Fri\v cova 1, CZ-25165 Ond\v rejov, Czech Republic
         \and
             Earth and Planetary Science Department, Planetary Geosciences Institute, University of Tennessee, Knoxville, TN 37996, United States
     \and
         D\' epartement d'Astrophysique, G\' eophysique et Oc\' eanographie, Universit\' e de Li\`ege, All\`ee du Six Ao\^ ut 17, 4000 Li\`ege, Belgium
         \and
             Badlands Observatory, 12 Ash Street, P.O. Box 37, Quinn, SD 57775, USA
         \and
             Modra Observatory, Department of Astronomy, Physics of the Earth, and Meteorology, FMPI UK, Bratislava SK-84248, Slovakia
         \and
             Institute for Astronomy, University of Hawaii at Manoa, Honolulu, HI 96822, USA
         \and
             Instituto de Astrof\'{\i}sica de Canarias (IAC), C\/V\'{\i}a L\'{a}ctea s/n, 38205 La Laguna, Spain
         \and
             Departamento de Astrof\'{\i}sica, Universidad de La Laguna, 38206 La Laguna, Tenerife, Spain
         \and
         Planetary Science Institute, 1700 East Fort Lowell Road, Tucson, AZ 85719, USA
         \and
         446 Sycamore Ave., Eaton, CO  80615, USA
}

   \date{Received x-x-2016 / Accepted x-x-2016}
% \abstract{}{}{}{}{} 
% 5 {} token are mandatory
 
  \abstract
   {The near-Earth asteroid (3200)~Phaethon is an intriguing object: its perihelion is at only 0.14 au and is associated with the Geminid meteor stream.}
   {We aim to use all available disk-integrated optical data to derive a reliable convex shape model of Phaethon. By interpreting the available space- and ground-based thermal infrared data and Spitzer spectra using a thermophysical model, we also aim to further constrain its size, thermal inertia, and visible geometric albedo.}
   {We applied the convex inversion method to the new optical data obtained by six instruments and to previous observations. The convex shape model was then used as input for the thermophysical modeling. We also studied the long-term stability of Phaethon's orbit and spin axis with a numerical orbital and rotation-state integrator.}
   {We present a new convex shape model and rotational state of Phaethon: a sidereal rotation period of 3.603958(2) h and ecliptic coordinates of the preferred pole orientation of (319$^{\circ}$, $-$39$^{\circ}$) with a 5$^{\circ}$ uncertainty. Moreover, we derive its size ($D$=5.1$\pm$0.2~km), thermal inertia ($\Gamma$=600$\pm$200 \tiu), geometric visible albedo ($p_{\mathrm{V}}$=0.122$\pm$0.008), and estimate the macroscopic surface roughness. We also find that the Sun illumination at the perihelion passage during the past several thousand years is not connected to a specific area on the surface, which implies non-preferential heating.}
   {}
 
%   \keywords{minor planets, asteroids: general - photometry - models}

  \titlerunning{Characterization of near-Earth asteroid (3200) Phaethon}
  \maketitle

\section{Introduction}\label{sec:introduction}

%%%%%%%%%%%%%%%%%
% Send paper to Vishnu
% which lcs were obtained by Ron Dyvig?
%%%%%%%%%%%%%%%%%

The extraordinary near-Earth asteroid (3200)~Phaethon (hereafter Phaethon), which currently has a perihelion distance of only 0.14 au, regularly experiences surface temperatures of more than 1\,000 K. This B-type object is one of the best-studied low-perihelion asteroids \citep{Campins2009b}. 

%\footnote{Astronomical unit defined by the International Astronomical Union as 149\,597\,870\,700 meters.}

Phaethon has been dynamically associated with the Geminid meteor stream \citep[][and references therein]{Gustafson1989,Williams1993,Jenniskens2006}, one of the most prominent annually periodic meteor streams. For this reason, activity around Phaethon has been searched for. However, detecting the activity has always been challenging, in particular near perihelion, because of its close approaches to the Sun. \citet{Jewitt2010} and \citet{Li2013} succeeded using STEREO (a solar observatory) spacecraft data from 2009 and 2011 to measure near-Sun brightening of Phaethon by a factor of two, associated with dust particles of an effective diameter $\sim$1~$\mu$m ejected from the surface. On the other hand, there is no evidence of gas release \citep{Chamberlin1996}. 
However, the observed dust produced during the perihelion passage \citep[mass of $\sim$3$\times$10$^5$~kg,][]{Jewitt2013} is small compared to the mass of the whole Geminid stream \citep[mass of $\sim$10$^{12}$--10$^{13}$~kg,][]{Jenniskens1994}. Moreover, these small particles are quickly swept away by the solar radiation pressure and can hardly be reconciled with the age of the Geminids. 
The Geminid stream consists of much larger particles than those estimated from the STEREO data, with sizes from 10~$\mu$m to about 4--4.5 cm \citep{Arendt2014,Yanagisawa2008}. Some of them might even survive the passage through the Earth's atmosphere \citep{Madiedo2013} and drop meteorites that have never been collected. The mechanism(s) of mass loss from Phaethon, capable of producing the massive swarm of the Geminids, is not fully understood \citep{Jewitt2015}. So far, the most convincing process consists of thermal disintegration of the asteroid surface \citep[e.g.,][]{Delbo2014} assisted by rotation and radiation pressure sweeping, of thermal desiccation cracking, or both \citep[see the review of][]{Jewitt2015}. However, to reliably explain the total mass of the Geminid stream in the scope of its expected dispersion age of $\sim$10$^3$ years \citep{Ohtsuka2006}, additional theoretical and laboratory constraints of the thermal fracture mechanism as well as physical and surface properties of Phaethon are necessary.

For the physical properties of Phaethon, \citet{Licandro2007} found that the spectral shape of Phaethon is similar to that of aqueously altered CI/CM meteorites and of hydrated minerals. However, it is unclear whether the heating occurs as a result
of the close approaches to the Sun or in Phaethon's parent body (or both). The large main-belt asteroid (2) Pallas has been identified \citep{deLeon2010} from spectroscopic and dynamical arguments as the source of Phaethon. However, the spectrum of Phaethon is generally bluer than that of Pallas. Whether this spectral slope difference is due to the extreme solar heating on the former compared to the latter or to different grain sizes of the regolith of these bodies is still unclear \citep{deLeon2010}. If the spin state had an obliquity of $\sim$90$^{\circ}$ \citep{Krugly2002, Ansdell2014}, one of the hemispheres of Phaethon would receive substantially more solar heating during the perihelion passage because it is always facing the Sun. But \citet{Ohtsuka2009} reported no significant spectral variability across the body, which might indicate that, contrary to expectations \citep{Hiroi1996}, the sun-driven heating does not affect the spectral properties.
Alternatively, Phaethon may not always be exposing the same hemisphere at perihelion, or the reason for the obliquity might be a combination of both explanations. Interestingly, the composition of Pallas, deduced from spectroscopic data near three microns, matches those of heated CM chondrites and is also similar to that of the CR chondrite Renazzo \citep{Sato1997}. This is a result of the spectroscopic signature of phylosilicates. Even more interestingly, \citet{Madiedo2013} also indicated that the composition of the Geminids, deduced from spectra of atmospheric flashes, is consistent with those of CM chondrites. On the other hand, a good match of Phaethon's near-IR spectra with those of CK chondrites was also found \citep{Clark2010b}.

For the orbital evolution of Phaethon, we note that \citet{Chertenenko2010}, and also \citet{Galushina2015}, reported to have detected a transverse acceleration component in Phaethon's heliocentric motion. \citet{VokrouhlickyAIV2015} obtained a similar result, but their uncertainty was larger than that of \citet{Chertenenko2010}, such that the signal-to-noise ratio was only about $\simeq 1.4$, and for this reason the value was not listed in their Table~1 (we do not know the reason for the difference). We have made use of our pole solutions from Sect.~\ref{sec:shapeModel} and verified that the magnitude of the transverse acceleration reported by \citet{Chertenenko2010} is compatible with predictions of the Yarkovsky effect, assuming low bulk density $\simeq 1$~g~cm$^{-3}$ and thermal inertia
of $\simeq 600$ \tiu or slightly lower (Sec.~\ref{sec:TPMresults}). Alternately, the recoil effects of outgassing and particle ejection near perihelion may also contribute to the effective transverse orbital acceleration. \citet{Galushina2015}, however, estimated that the Yarkovsky effect should be about an order of magnitude stronger. The current situation suggests that we are on the brink of determining the Yarkovsky effect for Phaethon, and it is very reasonable to expect that it will be fairly well constrained during its upcoming close encounter in December 2017 (especially if radar astrometry can be obtained). Moreover, Gaia astrometry will also significantly help to constrain the Yarkovsky effect. While certainly more photometric and thermal observations will be taken late in 2017, it would be interesting to collect the current knowledge about Phaethon's physical parameters relevant for determining the Yarkovsky effect (such as the size, albedo, thermal inertia, spin axis direction, macroscopic shape, and surface roughness). If a sufficiently rich dataset is available, the Yarkovsky effect allows determining the asteroid bulk density: see \citet{VokrouhlickyAIV2015} for general overview, and \citet{Chesley2014}, \citet{Emery2014}, \citet{Rozitis2013}, \citet{Rozitis2014b} and \citet{Rozitis2014} for specific cases. We note that knowing the bulk density of asteroids is fundamental to shed light on their internal structure (such as monolithic vs rubble pile). For example, when the bulk is compared with the densities of meteorites, the porosity of asteroid interiors can be deduced. These physical properties of asteroids reflect the accretional and collisional environment of the early solar system. 

%In any case, we point out that the astrometry of \citet{Galushina2015} with a negative sign of da/dt, would  matches well an obliquity $>$90 deg, as it is the case for many other near-Earth asteroids \citep{LaSpina2004}. 
% update the LASPINA ET AL.
%For our purposes, however, it is not critical not to include this effect in the estimate of the past orbit evolution, as will be noted below.

The spin state of Phaethon was first constrained by \citet{Krugly2002}: they derived two pole solutions with low ecliptic latitude (about --10 deg). The recent work of \citet{Ansdell2014} based on additional photometric data reported a single pole solution compatible with one of the pole solutions of \citet{Krugly2002}. This provides a convex shape. Ansdell et al.'s shape model is based on disk-integrated optical data and computed with the convex inversion method \citep{Kaasalainen2001a,Kaasalainen2001b}. However, from their Fig.~3 it is obvious that the rotation period is not unique and their reported interval includes several local minima that correspond, in principle, to different pole solutions (evident in their Figs.~4~and~5). We are therefore convinced that the shape model needs additional attention before using it in further applications. To do this, we refined the shape model by using new optical data.

Thermal inertia $\Gamma$, size $D$, Bond albedo $A$, and surface roughness can be derived by using a thermophysical model \citep[hereafter TPM,][]{Lagerros1996, Lagerros1997, Lagerros1998} to analyze thermal infrared data \citep[see, e.g.,][for a review]{DelboAIV2015}. Although thermal infrared data of Phaethon exist that allowed determining a radiometric diameter \citep{Green1985,Tedesco2002,Usui2011,Usui2013}, there is currently no estimate available for the thermal inertia for this body. 

In Sect.~\ref{sec:data} we describe optical data that we used for the shape modeling and the thermal infrared data and Spitzer spectra that made the thermophysical modeling possible. Light-curve inversion and TPM methods are presented in Sect.~\ref{sec:methods}. We derive a convex shape model of Phaethon in Sect.~\ref{sec:shapeModel} and use it in the TPM in Sect.~\ref{sec:TPMresults}. The orbital and spin axis evolution is discussed in Sect.~\ref{sec:dynamics}. Finally, we conclude our work in Sect.~\ref{sec:conclusions}.

\section{Data}\label{sec:data}

\subsection{Optical disk-integrated photometry}\label{sec:photometry}

We gathered a total of 55 dense-in-time light curves of Phaethon spanning 1994--2015, including 15 light curves from \citet{Ansdell2014}, 3 light curves from \citet{Pravec1998}, one light curve from \citet{Wisniewski1997}, and 7 light curves from \citet{Warner2015}. In addition, we obtained 29 new light curves with six different instruments. All light curves are listed in Table~\ref{tab:photometry}.

All light curves are based on aperture photometry in standard filter systems, either differential or absolute. The images were bias- and flat-field corrected, mainly using sky flats. As the data were obtained by different telescopes and by many different observers, the photometry reduction procedures may vary slightly, but all follow the standard procedures. Some of the data were initially absolutely calibrated, but we normalized them like all remaining light curves. Only the relative change of the brightness due to rotation and orientation with respect to the Sun and the observer is necessary for light-curve inversion. Moreover, the epochs were light-time corrected.

We obtained four light curves of Phaethon with the University of Hawaii 2.2-meter telescope (UH88) located near the summit of Maunakea in Hawaii between August and October 2015. We used the Tektronix 2048x2048 CCD camera, which has a ${7.5}'\mathrm{x}{7.5}'$ field of view corresponding to a pixel scale of ${0.22}''$. The images obtained were nyquist-sampled corresponding to the typical seeing of $\sim {0.8}'$ during the observations and to reduce readout time. Exposures were between 120--180 s in the Sloan ${r}'$ filter. Non-sideral tracking at half the rate of Phaethon was used so that the PSFs of the asteroid and background stars have similar morphologies. The light curves taken with the UH88 are semi-dense with 15-25 minutes between observations and have sufficient density because the rotation period of Phaethon is 3.6 h. Semi-dense light curves were obtained to allow the simultaneous observations of additional targets.

We also used the Centre P\'edagogique Plan\` ete et Univers
(C2PU) 1.04 meter telescope situated in the Calern observing station of the Observatoire de la C\^ ote d'Azur  in France (06$^{\circ}$ 55' 22.94'' E, 43$^{\circ}$ 45' 13.38'' N, 1270 m, IAU code 010). This telescope has a f/3.2 prime focus with a three-lenses Wynne corrector and a QSI632 CCD camera with a built-in filter wheel. We obtained seven light curves between December 2014 and February 2015.

Four light curves from 2004 and 2007 were obtained at Modra observatory in Slovakia \citep{Galad2007}. Open-filter and standard aperture photometry were used.

We used the 66 cm f/4.8 Newtonian telescope at Badlands Observatory, Quinn, South Dakota, to observe Phaethon over the course of two weeks ending on December 12, 2004. The data were obtained using an Apogee 1Kx1K CCD camera with SiTe detector. After basic calibration, the data were reduced using Canopus \citep{Warner2006e}.

Data from apparitions in 1994, 2003, and 2004 consisting of eight light curves were obtained by the 65 cm telescope in Ond\v rejov, Czech Republic. In all cases, Cousins R filter and aperture photometry were used. Moreover, data from 2004 were absolutely calibrated in the Cousins R system with \citet{Landolt1992} standard stars with absolute errors of 0.01 mag.

Four light curves from the apparition in 1998 were obtained with the 82cm IAC-80 telescope at Teide Observatory (Canary Islands, Spain) using a broadband Kron-Cousins R filter. We used standard aperture photometric procedures and performed absolute photometry using at least three Landolt field stars \citep{Landolt1992}. The exposure time was 300 seconds during all nights, and because
of the trailed stars, only absolute photometry was possible. A Thomson 1024$\times$1024 CCD chip was used, offering a field of nearly 7.5 arcmin.

The data from UH88 and Ond\v rejov were reduced with our custom-made aperture photometry software Aphot+Redlink developed by Petr Pravec and Miroslav Velen.

%Moreover, we also used 108 sparse-in-time measurements from Catalina Sky Survey \citep{Larson2003} and the Lowell \citep{Bowell2014a} re-calibrated sparse photometry that was originally submitted to the Minor Planet Center (472 single measurements). Data from both sources were transformed from magnitudes to fluxes, the fluxes were normalized to the one au distance from the Sun, the epochs were light-time corrected and clear outliers were rejected \citep[for details see][]{Hanus2011}. Note that data from Catalina are a subset of Lowell data, thus both datasets should be treated independently.

Our light curves in the standard format used for the light-curve inversion (i.e., epochs and brightness are accompanied by ecliptic coordinates of the Sun and Earth centered on the asteroid) are available in the Database of Asteroid Models from Inversion Techniques \citep[DAMIT\footnote{\texttt{http://astro.troja.mff.cuni.cz/projects/asteroids3D}},][]{Durech2010}.

\subsection{Thermal infrared data}\label{sec:IR}

\begin{figure}
    \begin{center}
        \resizebox{1.0\hsize}{!}{\includegraphics{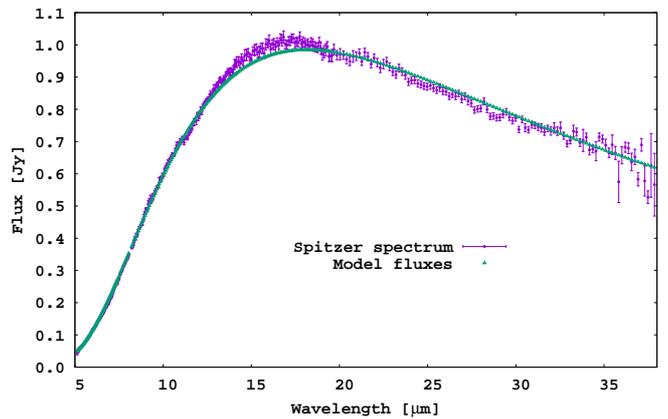}}\\
        %\resizebox{1.0\hsize}{!}{\includegraphics{fig/3200_emissivity_1.eps}}\\
    \end{center}
    \caption{\label{fig:spectra}Spitzer IRS spectral data of Phaethon from January 14, 2005. We show the observed spectra and the best-fitting TPM model for the first pole solution ($D=5.1$ km, $p_{\mathrm{V}}=0.122$, $\Gamma=600$~\tiu, high macroscopic roughness).}
\end{figure}

Measurements of asteroids in thermal (mid-) infrared are, in general, difficult to obtain, therefore it is not surprising that only a few previous measurements are available for Phaethon. The IRAS satellite observed Phaethon in 1983 in four filters, providing a total of 19 individual measurements at six epochs (see Table~\ref{tab:IR}). We extracted IRAS space-based observations of Phaethon from the SIMPS database of \citet{Tedesco2002}. Each epoch consists of thermal infrared data in filters with isophotal wavelengths at 12, 25, and 60~$\mu$m. Moreover, one epoch contains flux at 100~$\mu$m as well. These data were previously analyzed by means of a standard thermal model: a radiometric diameter of $5.1\pm0.2$~km and a geometric visible albedo of $0.11\pm0.01$ were derived by \citet{Tedesco2004}.

We also extracted ground-based observations presented by \citet{Green1985}, which resulted in 12 measurements at nine different wavelengths. We excluded fluxes at wavelengths smaller than 4~$\mu$m, which were contaminated by the reflected-light component, and transformed fluxes into Jansky units (Tab.~\ref{tab:IR}). The authors reported a radiometric diameter of $4.7\pm0.5$~km and a geometric visible albedo of $0.11\pm0.02$.

The AKARI satellite \citep{Ishihara2010} observed Phaethon as well. Unfortunately, the fluxes are not publicly available. A
radiometric diameter ($4.17\pm0.13$~km) and a geometric visible albedo ($0.16\pm0.01$) were reported by \citet{Usui2011}.

These reported radiometric sizes are not fully compatible with each other. The differences are likely caused by the underestimation of model systematics (which means that the reported error bars are too small) and possible calibration errors.

Observations by the Spitzer InfraRed Spectrograph \citep[IRS][]{Houck2004} started at 22:14:54 on 14 January 2005 (UT) and were concluded
at about 22:24:30 UT on the same day. This spectrum covers mid-infrared wavelengths of 5--37 $\mu$m (Fig.~\ref{fig:spectra}). The Spitzer order-to-order flux uncertainties are 10\% or lower \citep{Decin2004}. We measured four orders and used an overall relative scaling for the final flux. We consider them as four independent absolute flux measurements, therefore we treat the final flux uncertainty as 5\% throughout, which should translate into 2.5\% in diameter. Typically, sizes derived by the thermophysical model have uncertainties of about 5\% or higher, which  means that the calibration uncertainty does not affect the final size uncertainties significantly when added in quadrature.

\section{Methods}\label{sec:methods}

\subsection{Light-curve inversion method}\label{sec:inversion}

As the typically non-spherical asteroids rotate around their rotational axis, they change the illuminated part of their bodies with respect to the observer, and thus exhibit temporal variations of their brightness. The light-curve inversion methods (LI) aim, under various assumptions, to search for unknown parameters (including sidereal rotation period, spin axis orientation, and shape) that affect the observed brightness. The most commonly used inversion method, which assumes a convex shape model, is the convex inversion of \citet{Kaasalainen2001a, Kaasalainen2001b}. 

We systematically ran this gradient-based LI method with different initial periods and pole orientations to sample the parameter space. A fine enough grid of initial parameter values guaranteed that we did not omit any local minimum. If the photometric dataset is rich in observing geometries, meaning that we sample all sides of the asteroid, we ideally derived one combination of parameters, that is, a solution that fits the observed data significantly better than all the other combinations (by means of a $\chi^2$ metric). To be more precise, we used a gradient-based method with the Levenberg-Marquardt implementation that converges to the closest minimum for a set of initial parameter values. We aim to find the global minimum in the parameter space by investigating all possible local minima. Moreover, our applied method assumes that the asteroid rotates along its principal axis with a maximum momentum of inertia. We checked throughout whether this condition was fulfilled for our final shape model.

\subsection{Thermophysical model}\label{sec:TPM}

A TPM calculates thermal infrared fluxes for a given set of physical parameters, illumination, and observing geometry of an asteroid. Classically, the shape and rotational state of the asteroid are considered as fixed inputs for the TPM. Such a TPM has been applied, for example, to shapes of asteroids (341843)~2008~EV$_5$ and (101955)~Bennu based on radar imaging \citep{AliLagoa2014, Emery2014}, and to convex shapes from optical data of asteroids (25143)~Itokawa and (1620)~Geographos \citep{Muller2014, Rozitis2014}. The downside of this approach is that it does not account for the uncertainties in the shape model and the rotation state. However, there is growing evidence that these uncertainties affect the TPM results \citep{Rozitis2014,Hanus2015a}. The recent TPM approach of \citet{Hanus2015a}, called the varied-shape TPM, is based on mapping the shape and rotation state uncertainties by generating various shape models of an asteroid from its bootstrapped optical data. These shape models are then used as inputs for the classical TPM scheme. %We will follow this approach in the analysis of the thermal infrared observations of Phaethon.

The TPM solves the heat-conduction equation over the relevant top surface layer of an airless body and computes temperatures for each surface element, usually a triangular facet. Thermal fluxes in desired wavelengths and directions can be then easily output as well. We used a TPM implementation of \citet{Delbo2007a} and \citet{Delbo2004} that is based on TPM developement by \citet{Lagerros1996, Lagerros1997, Lagerros1998}, \citet{Spencer1989}, \citet{Spencer1990}, and \citet{Emery1998}. This thermophysical model, recently used in \citet{AliLagoa2014} and \citet{Hanus2015a}, takes an asteroid shape model, its rotation state, and a number of physical parameters such as Bond albedo $A$, macroscopic surface roughness, and thermal inertia $\Gamma$ as input parameters. The macroscopic roughness is parametrized by an opening angle and areal density of a spherical crater on each surface element. This crater is divided into several
surface elements, typically 40, and a heat-conduction equation, accounting for shadowing and mutual heating, is solved in each one of these elements. We used five different combinations of the opening angle and areal density in the TPM and considered zero (opening angle=0$^{\circ}$, areal density=0), low (30$^{\circ}$, 0.3), medium (50$^{\circ}$, 0.5), high (70$^{\circ}$, 0.7) and extreme (90$^{\circ}$, 0.9) roughness. Knowledge of absolute magnitude $H$ and slope $G$ is required as well. 

By running TPM with different values of input parameters, and subsequently, by comparing computed fluxes with observed ones, we can constrain some or all of these parameters. Typically, we minimized the metric 

\begin{equation}\label{eq:chi2}
\chi^2 = \sum_i\frac{(s^2F_i-f_i)^2}{\sigma^2_i},
\end{equation}
where $f_i$ are the observed fluxes, $s^2F_i$ the modeled fluxes, where the scale factor $s$ corresponds to the asteroid size, and $\sigma_i$ represent the errors of fluxes $f_i$, where $i$ samples all individual measurements.

%As already mentioned, this TPM method does not account for the uncertainty in the shape model and the rotation state. Thus, we also used the varied-shape TPM approach of \citet{Hanus2015a} to validate the reliability of our solution based on the original shape and spin state solution.

\section{Results and discussions}\label{sec:results}

\subsection{Shape model and rotation state}\label{sec:shapeModel}

\begin{figure}
    \begin{center}
        \resizebox{1.0\hsize}{!}{\includegraphics{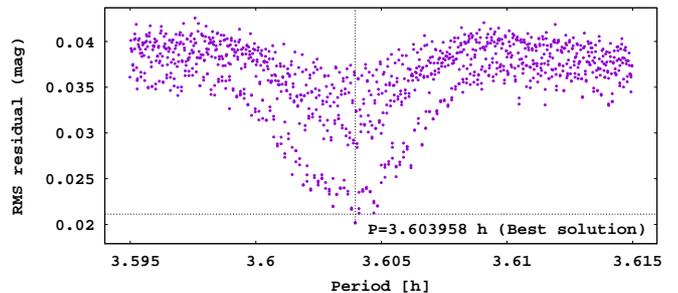}}\\
    \end{center}
    \caption{\label{fig:per}Light-curve inversion search for the sidereal rotation period of Phaethon: each point represents a local minimum in the parameter space (i.e., rotation period, pole orientation, and shape). The point with the lowest rms is the global minimum, and the horizontal line indicates a value with a 10\% higher $\chi^2$ than the best-fitting solution.}
\end{figure}

\begin{figure*}
    \begin{center}
        \resizebox{1.0\hsize}{!}{\includegraphics{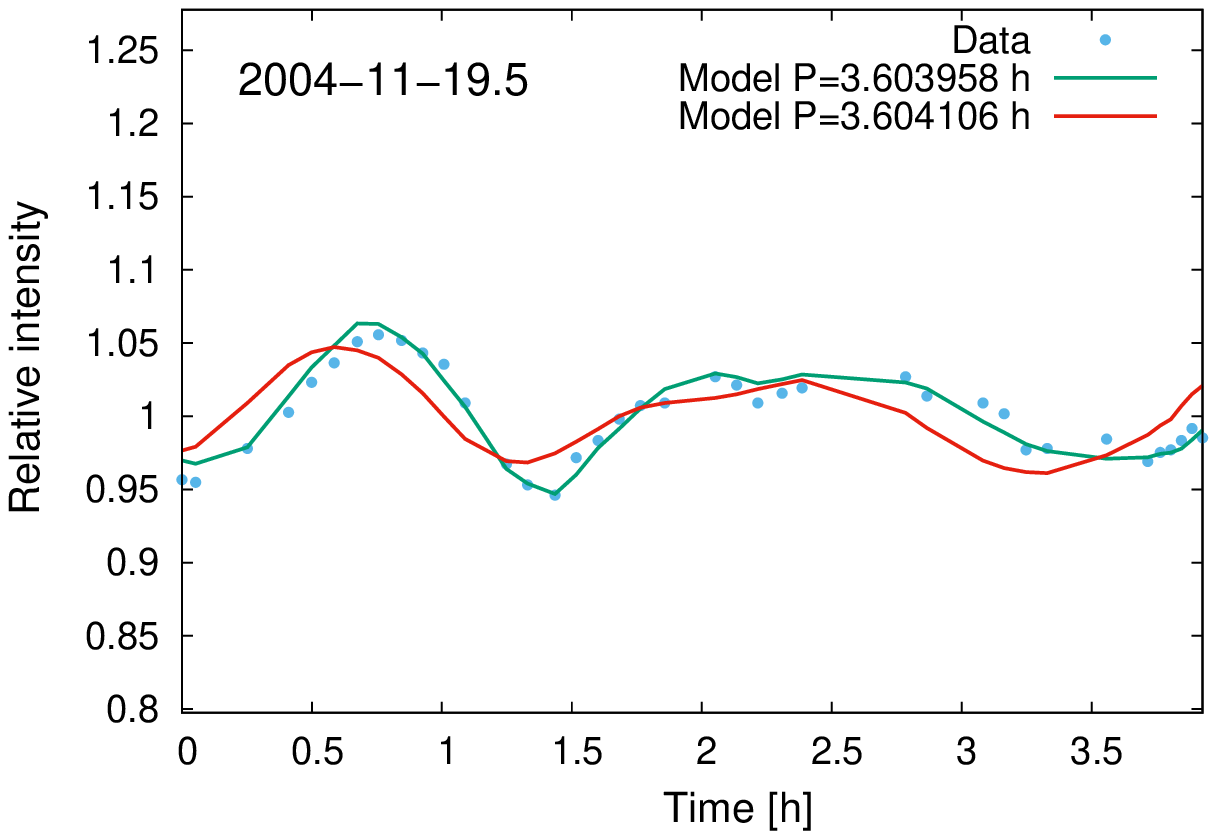}\includegraphics{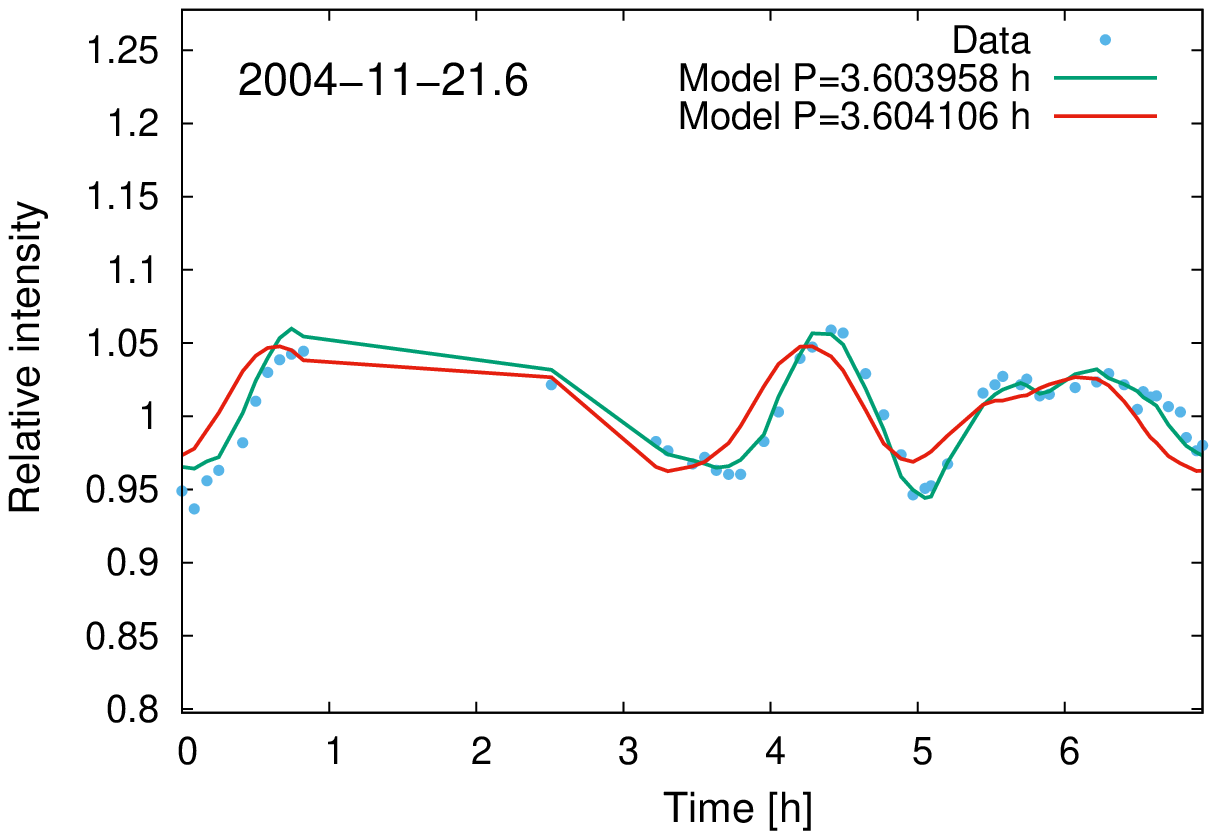}\includegraphics{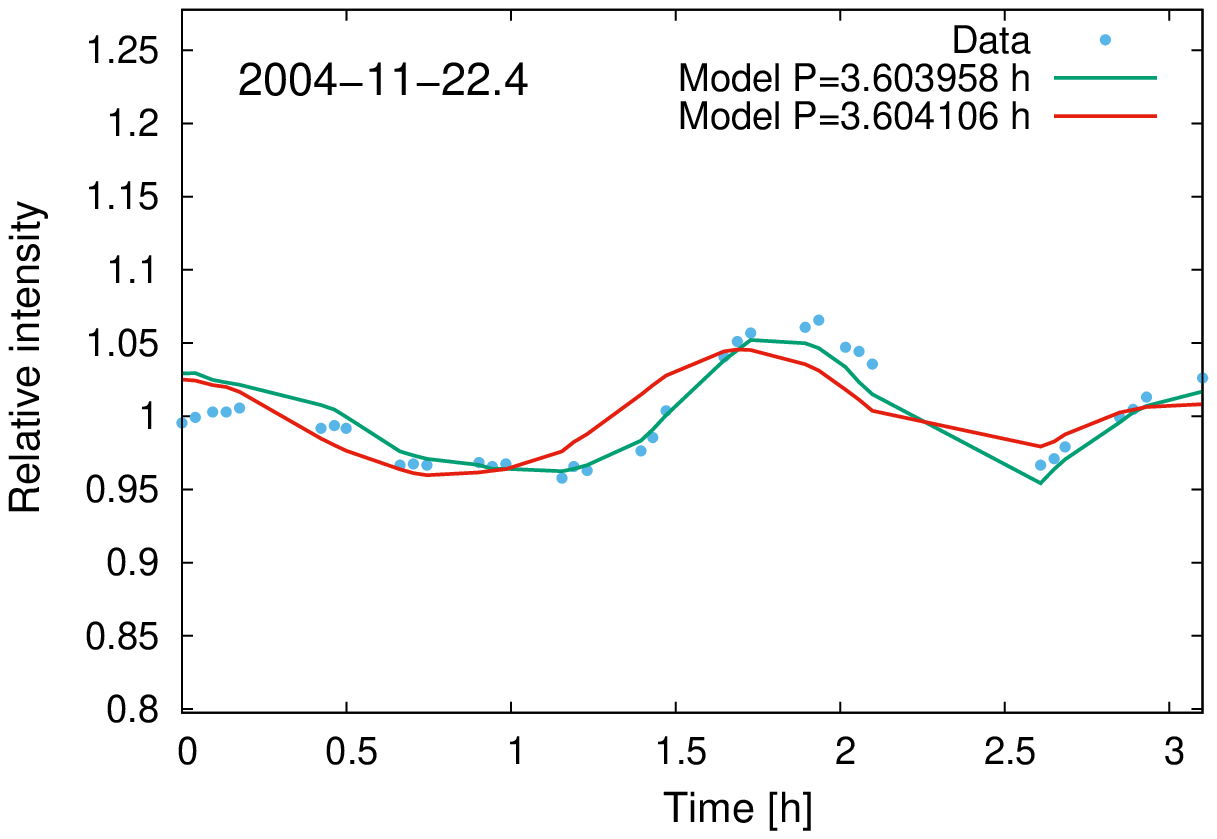}}\\
        \resizebox{1.0\hsize}{!}{\includegraphics{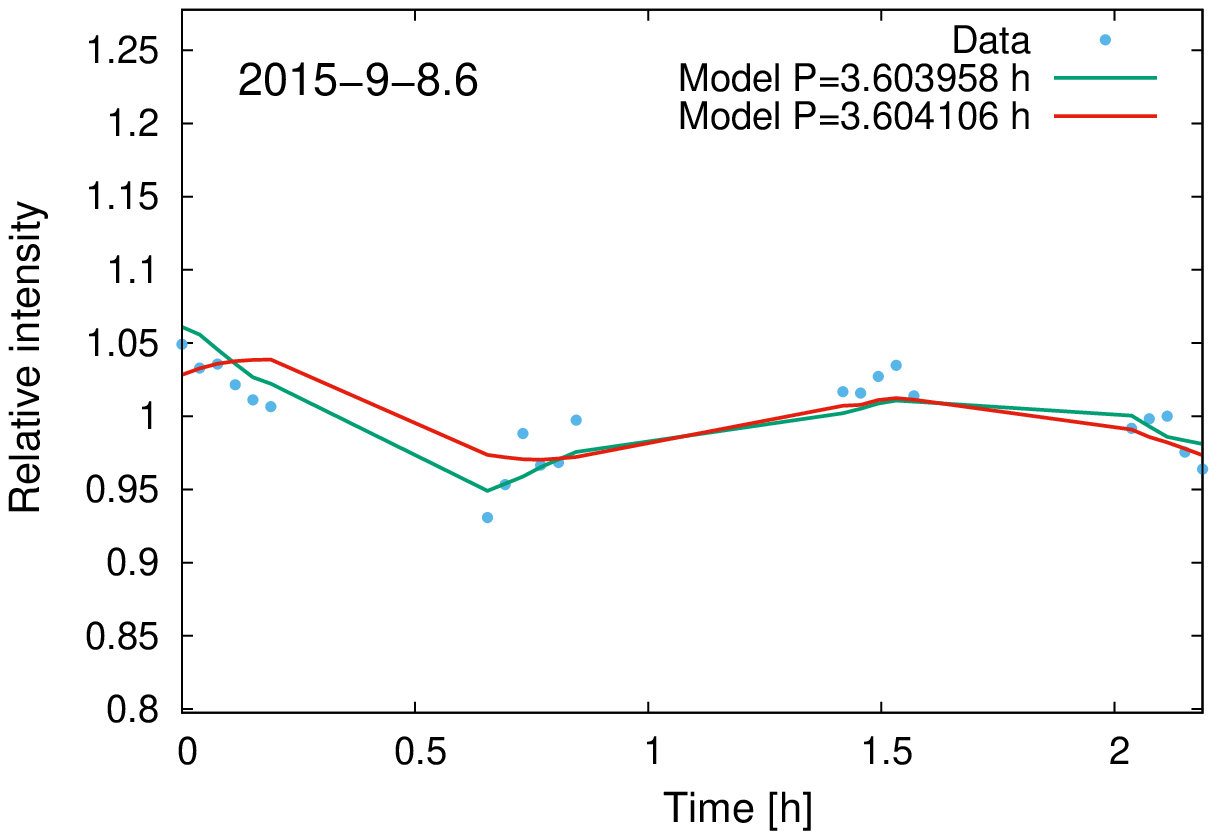}\includegraphics{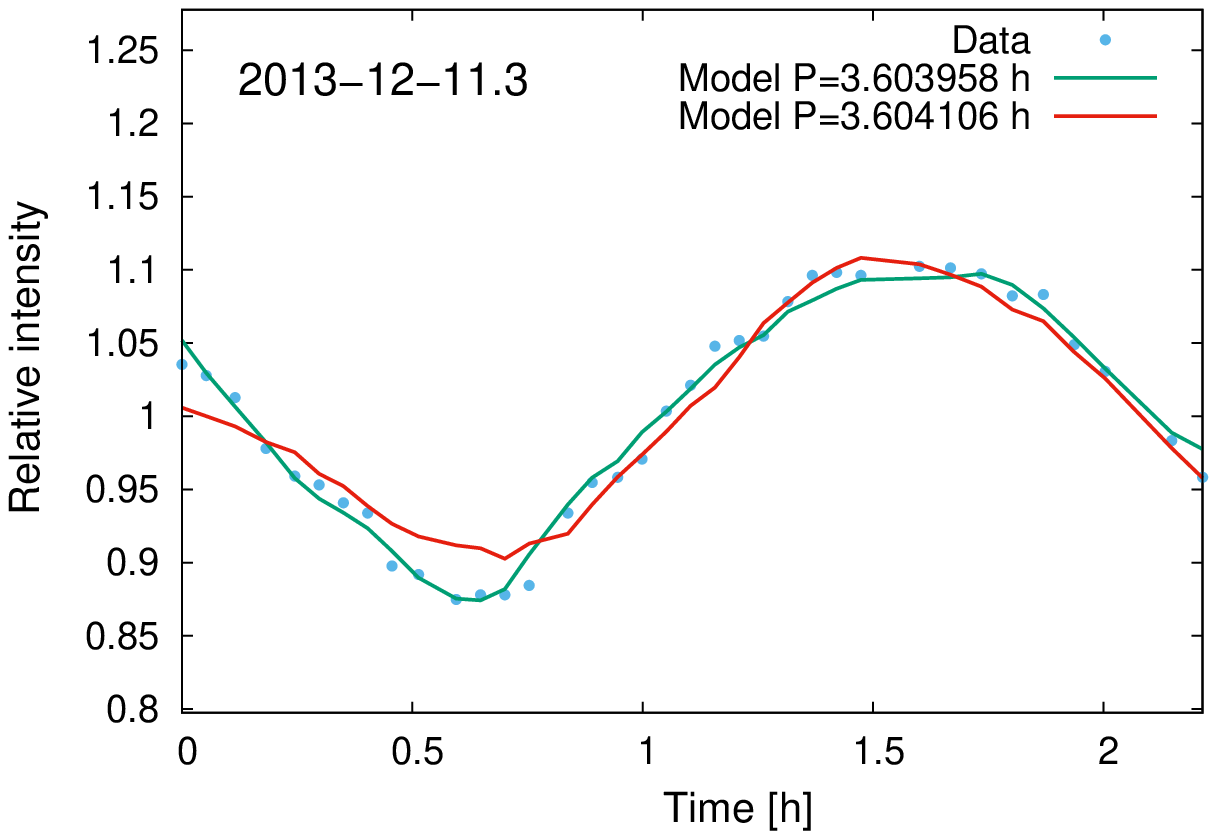}\includegraphics{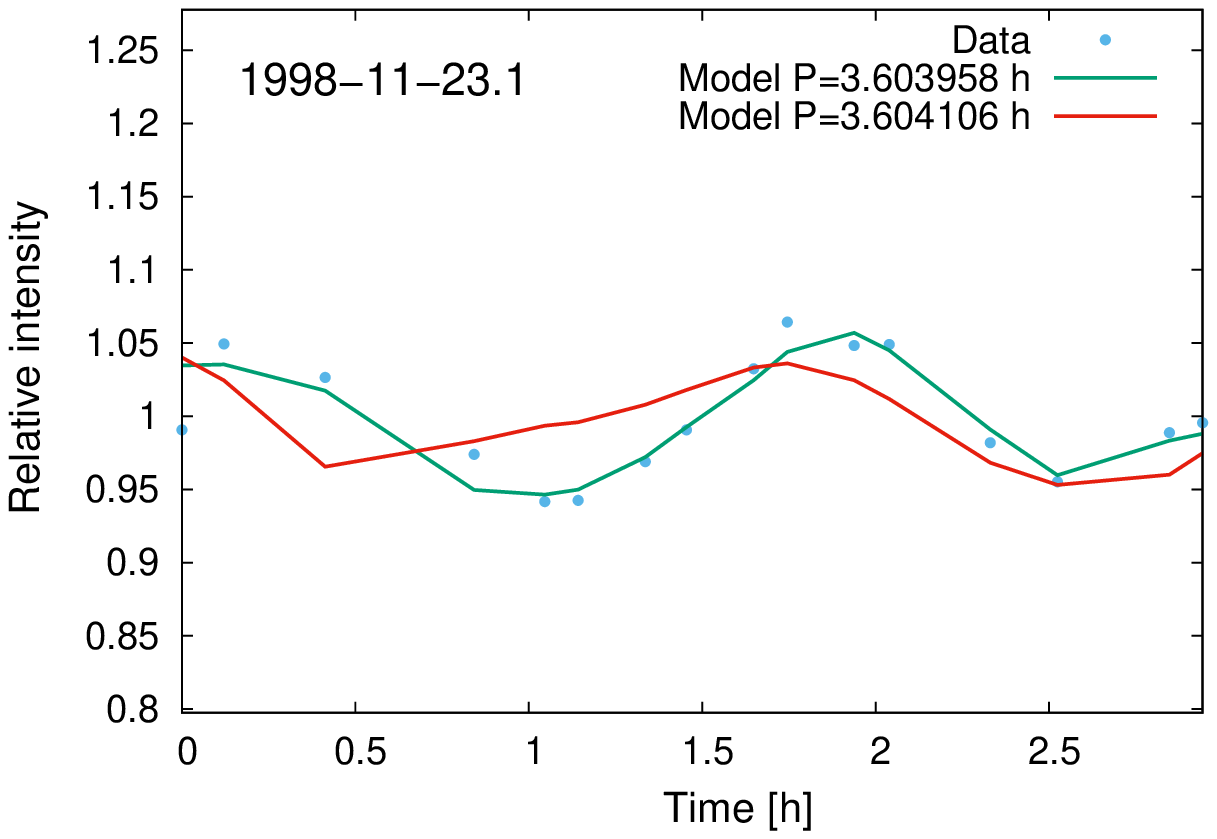}}\\
    \end{center}
    \caption{\label{fig:periods}Comparison between fits of several light curves that correspond to the best (green lines) and the second best (red lines) periods. The real measurements are plotted with blue dots.}
\end{figure*}

\begin{figure*}
    \begin{center}
        \resizebox{1.0\hsize}{!}{\includegraphics{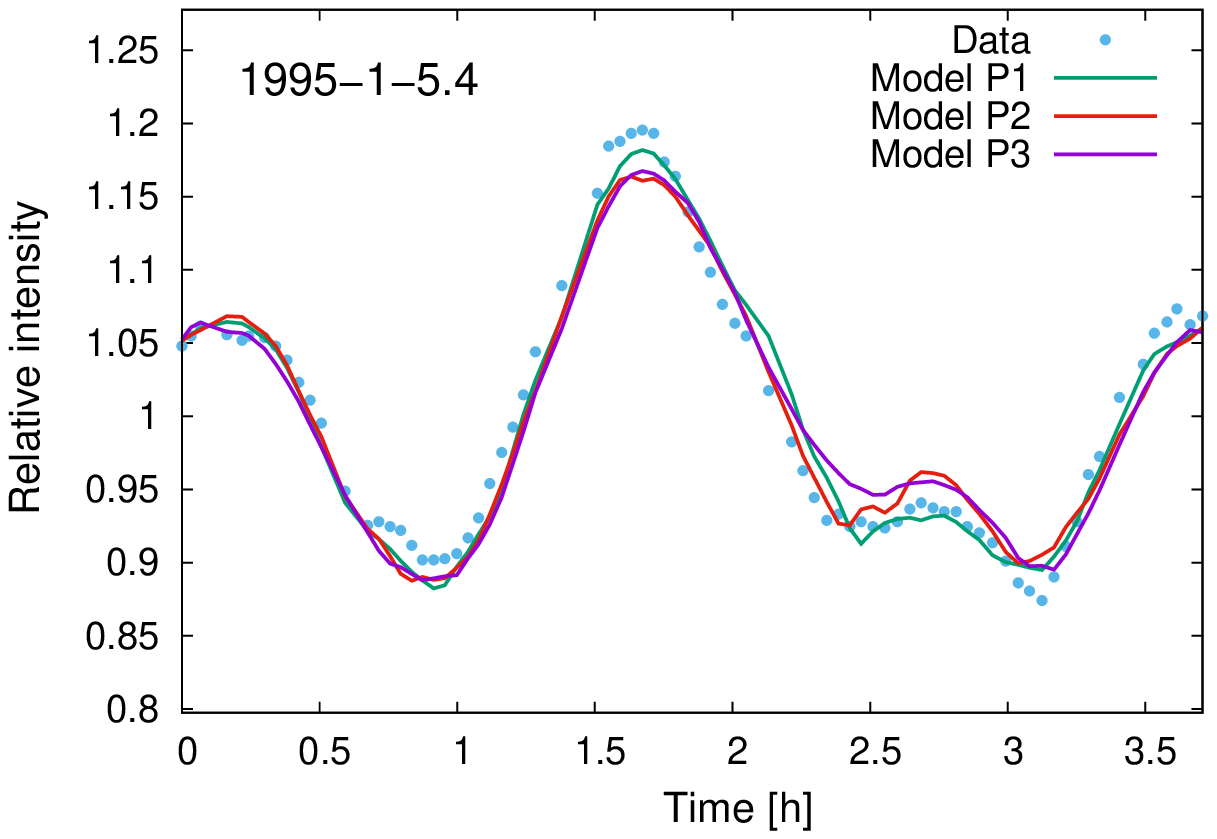}\includegraphics{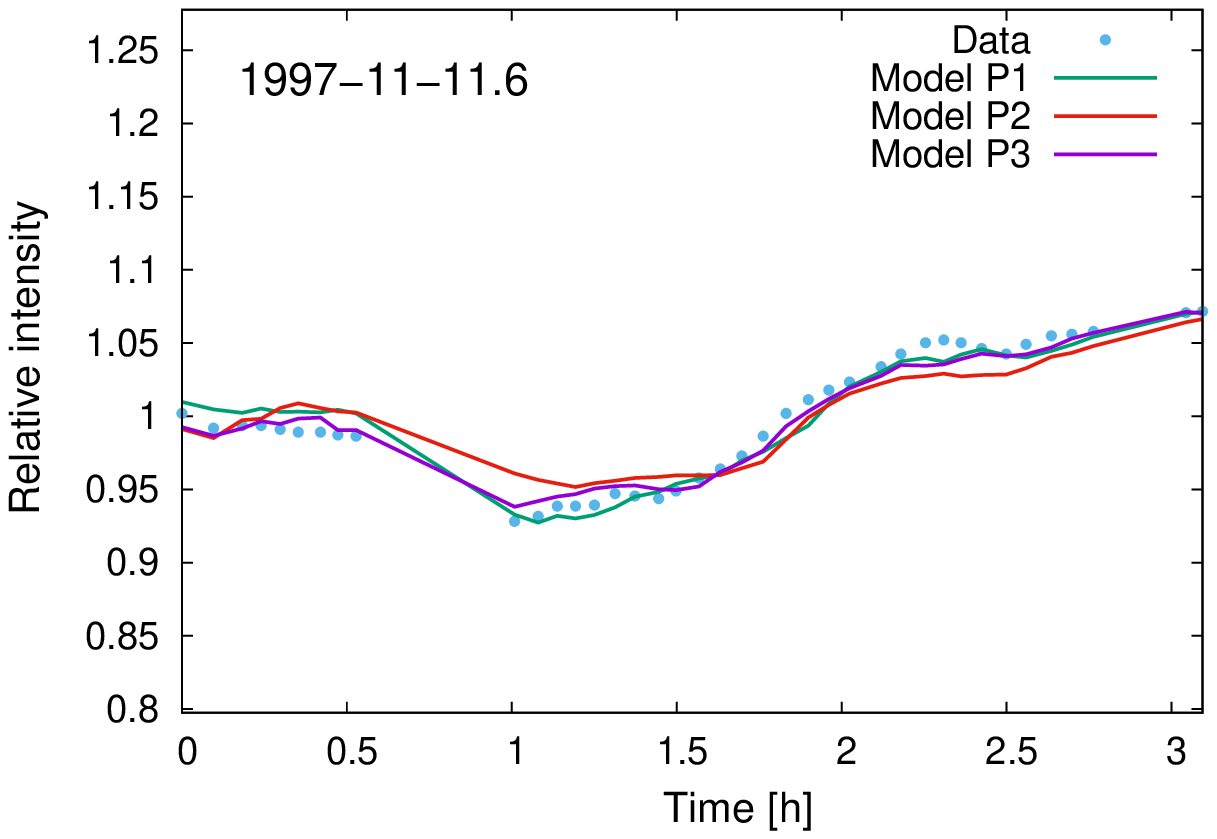}\includegraphics{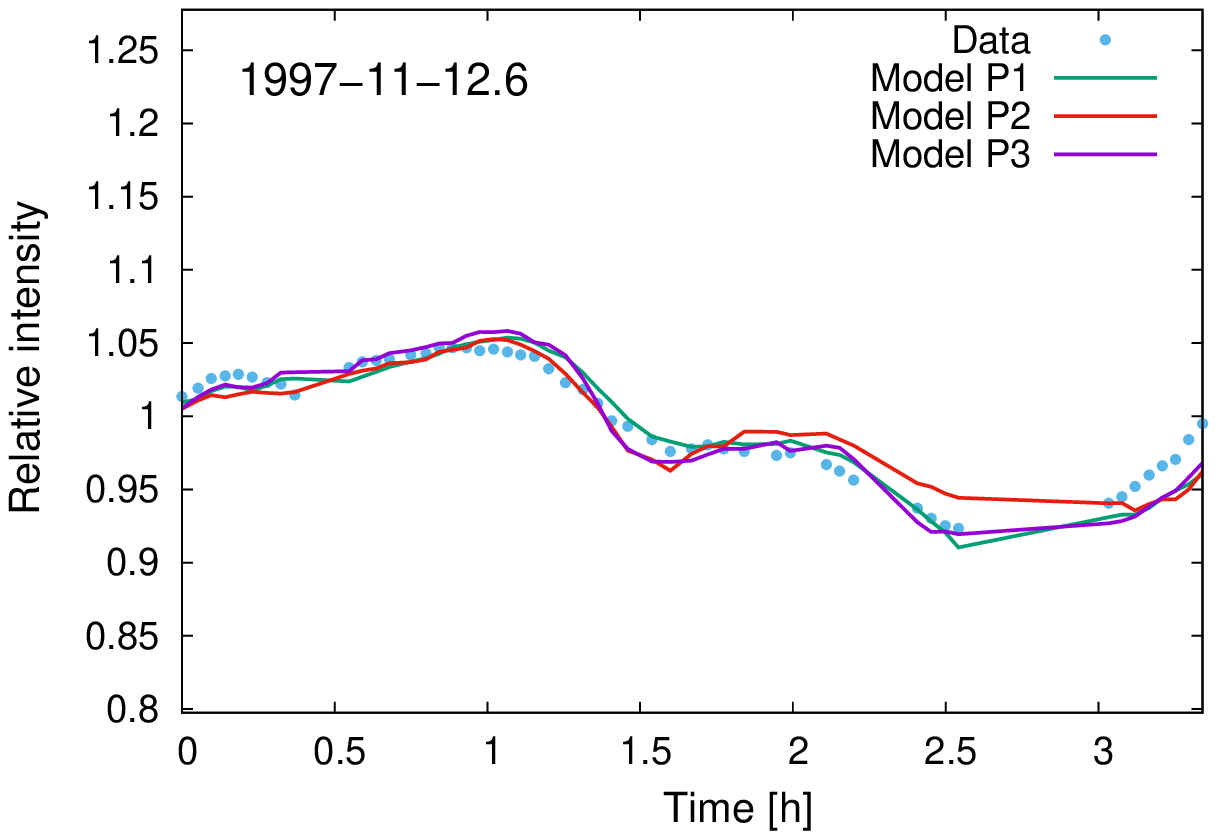}}\\
        \resizebox{1.0\hsize}{!}{\includegraphics{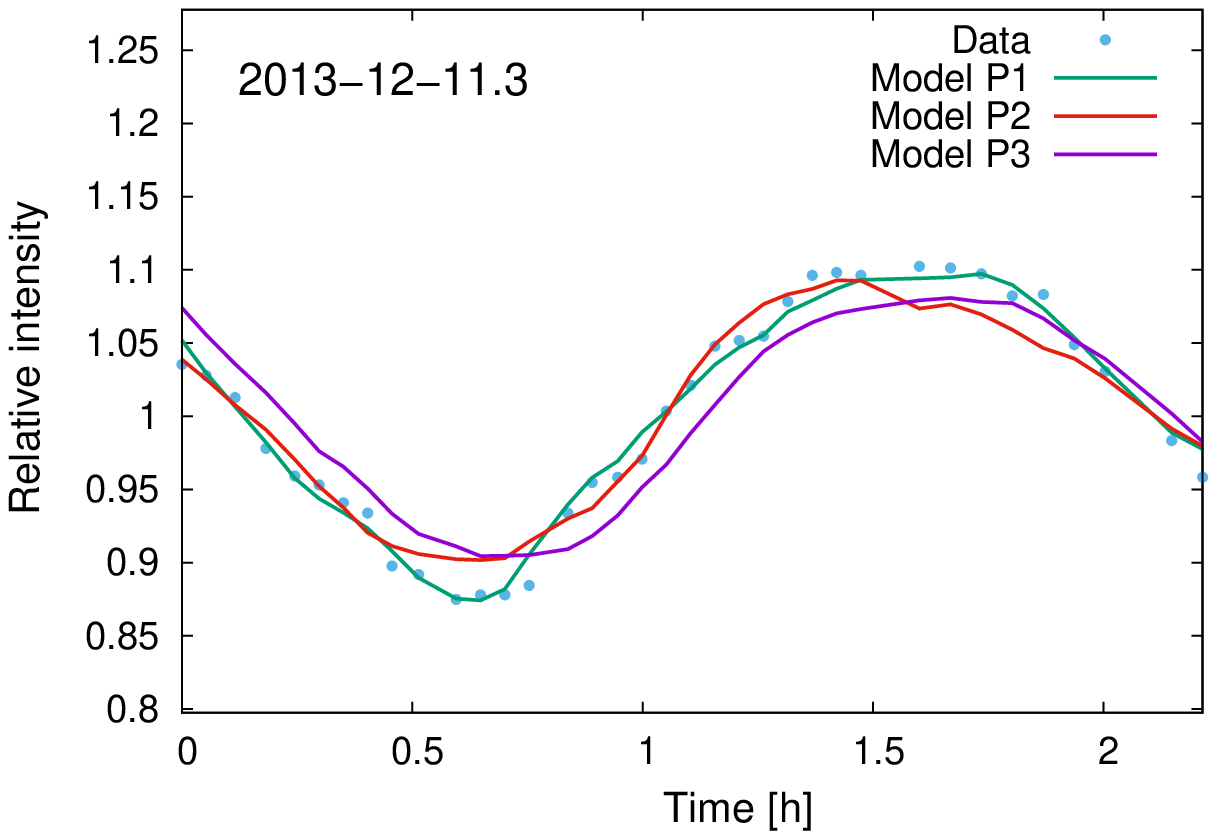}\includegraphics{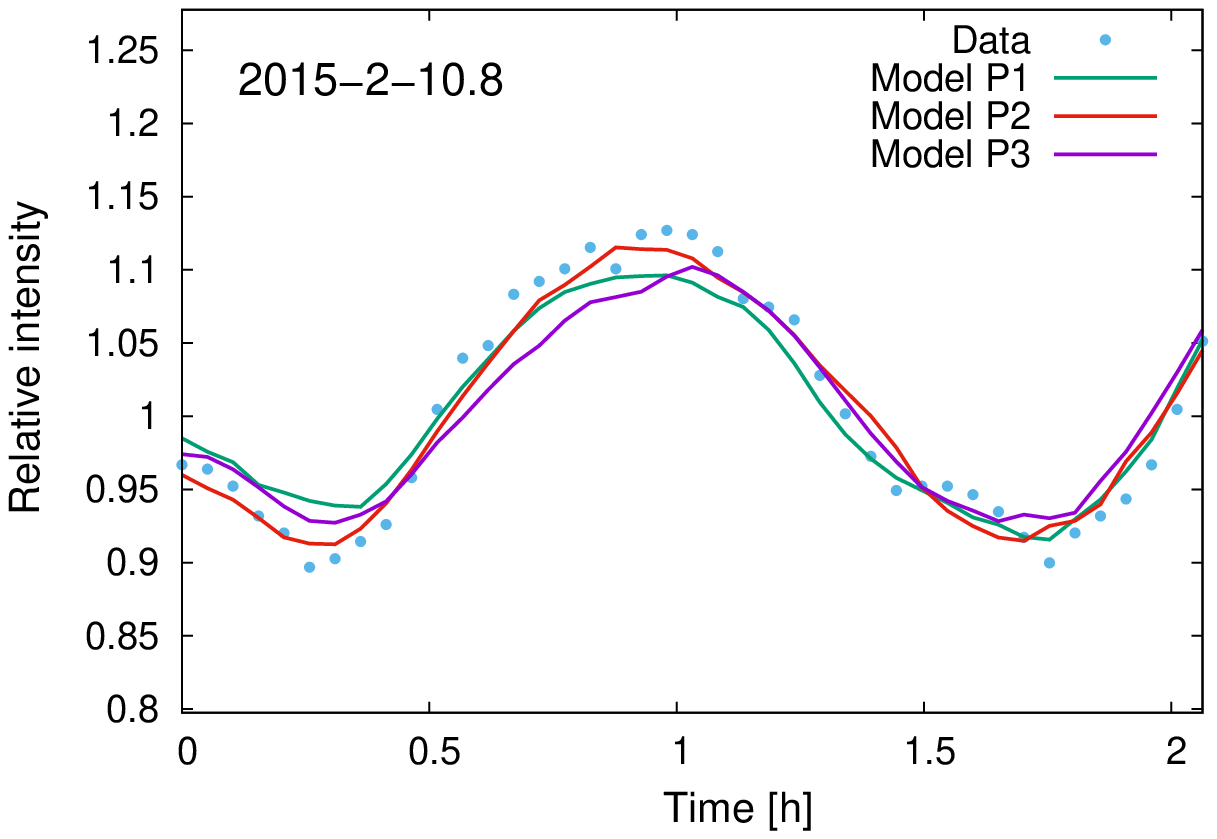}\includegraphics{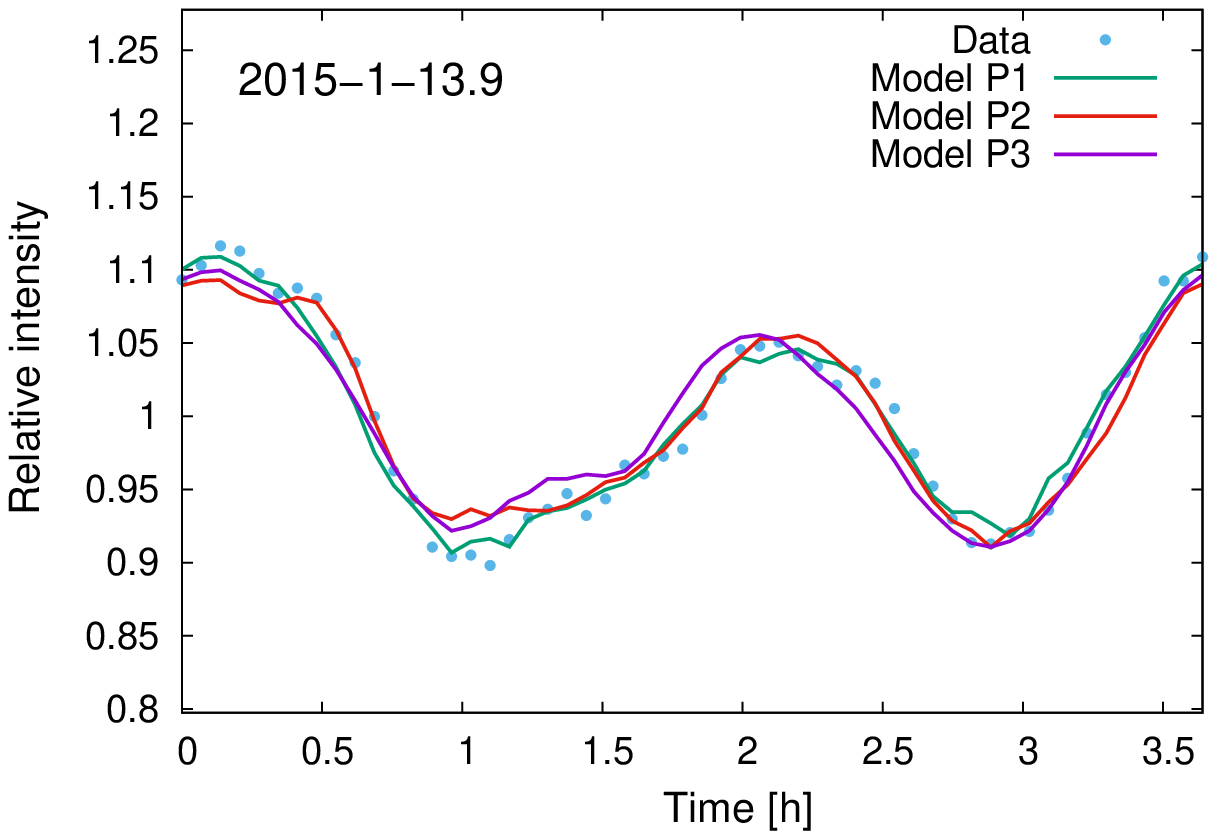}}\\
    \end{center}
    \caption{\label{fig:poles}Comparison between fits of several light curves that correspond to the best  (P1, green lines), the second best (P2, red lines), and the third best (P3, purple lines) pole solution. The real measurements are plotted with blue dots.}
\end{figure*}

\begin{figure}
    \begin{center}
        \resizebox{1.0\hsize}{!}{\includegraphics{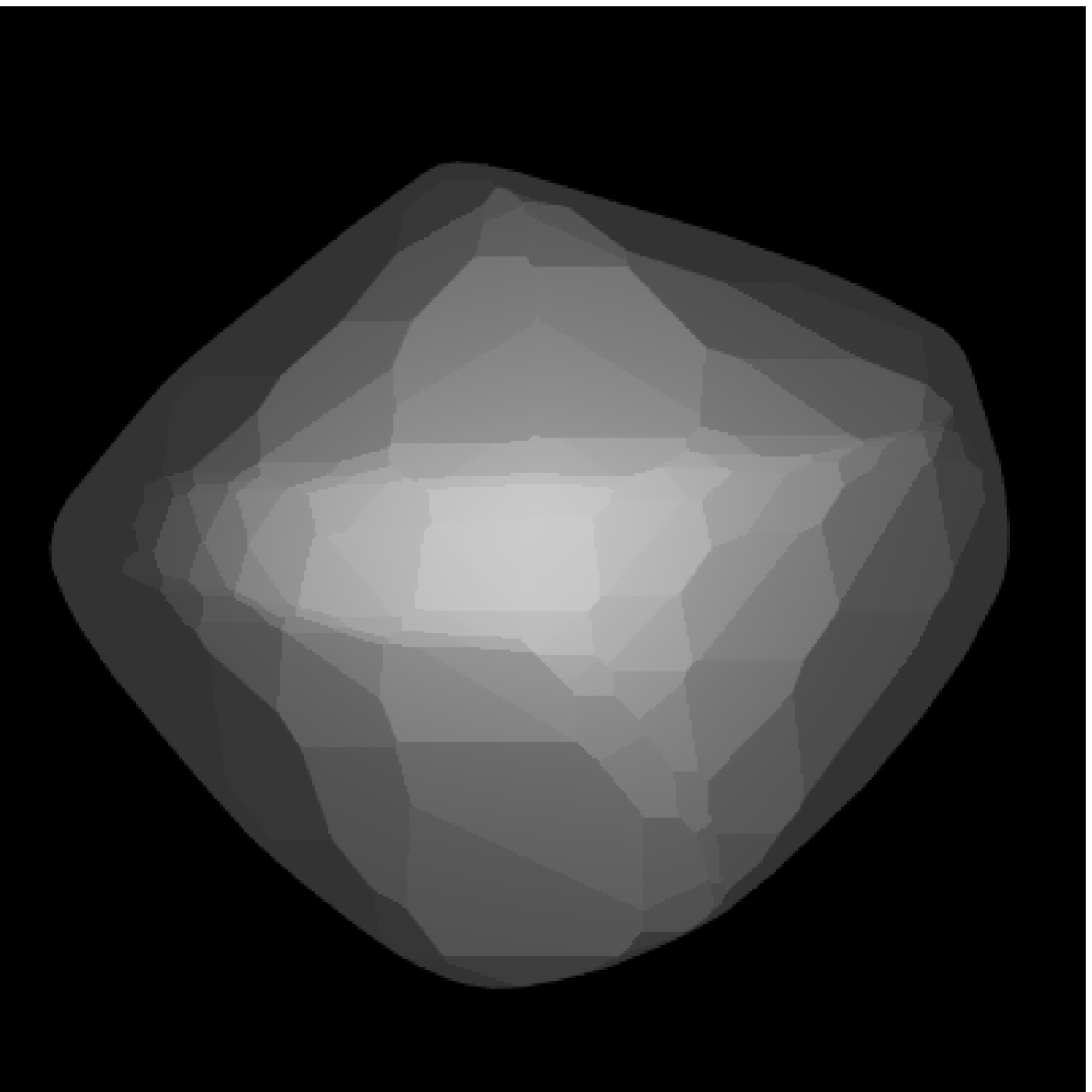}\includegraphics{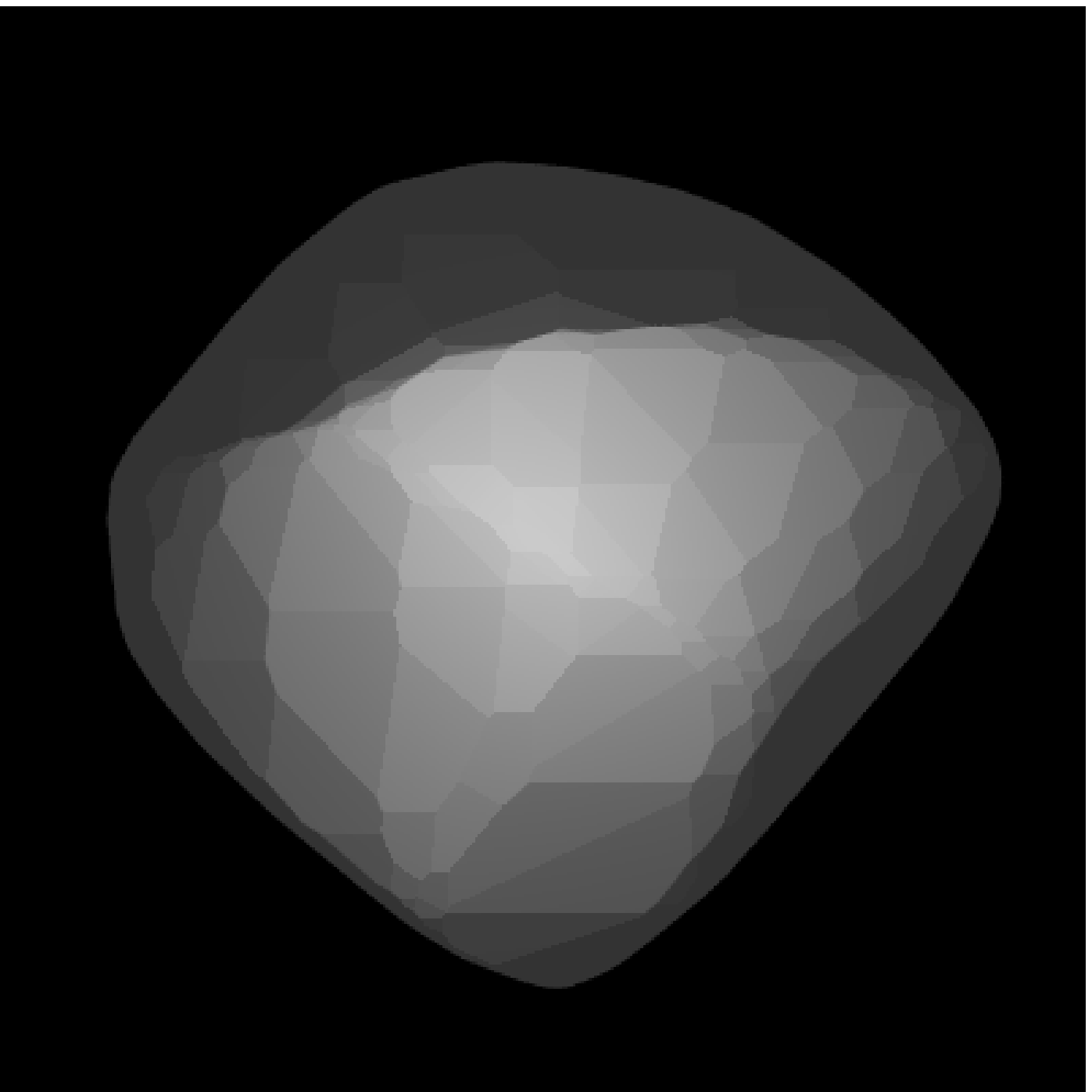}\includegraphics{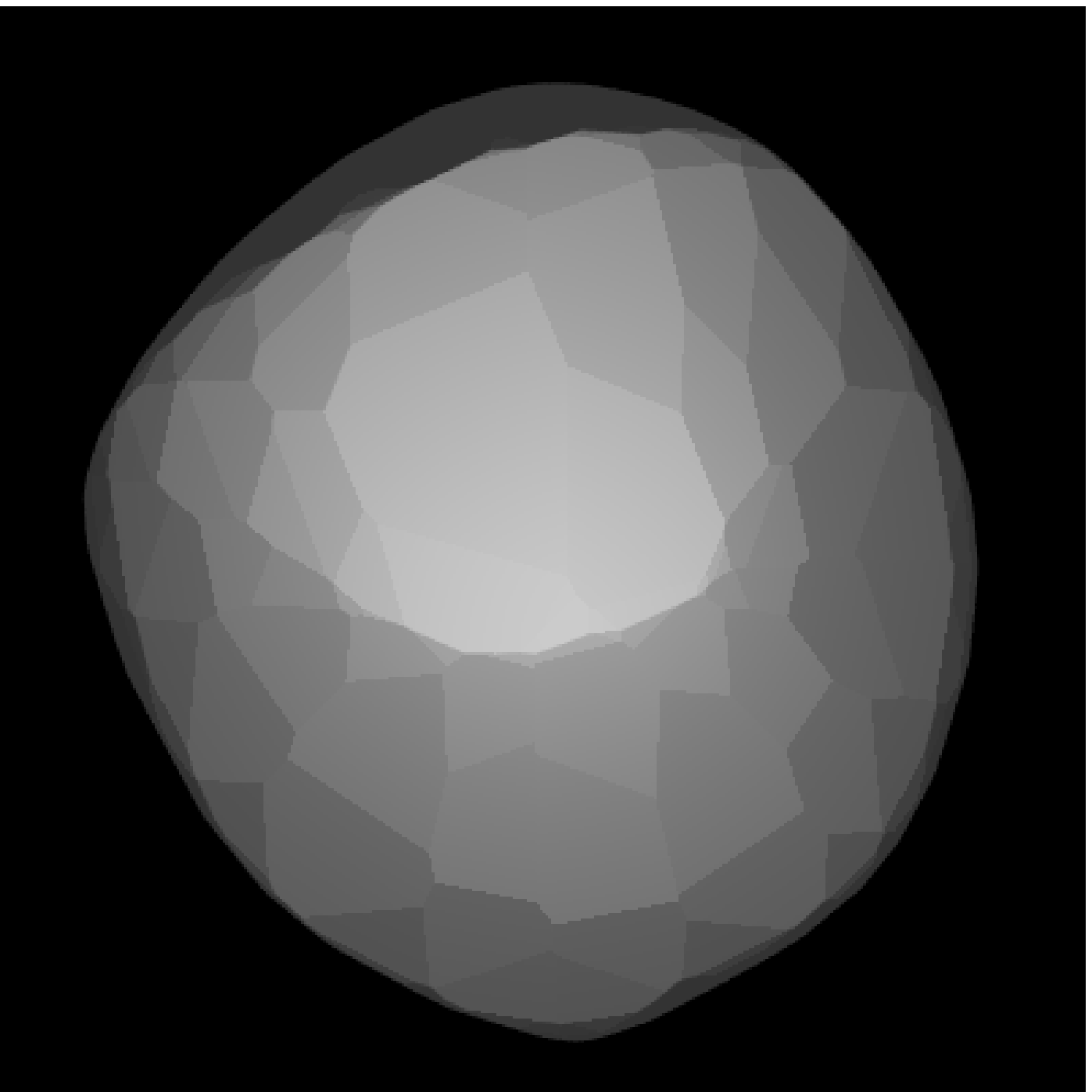}}\\
        \resizebox{1.0\hsize}{!}{\includegraphics{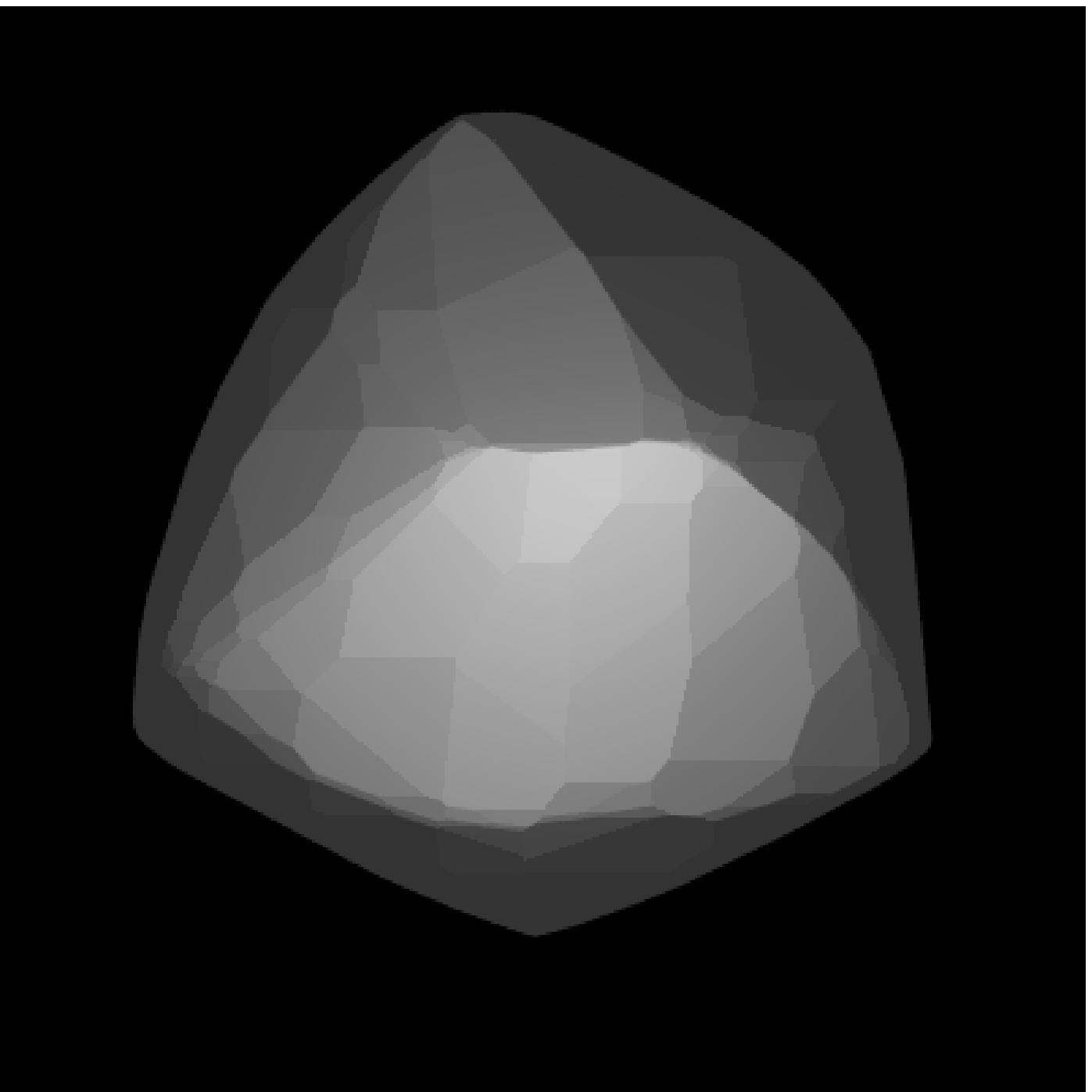}\includegraphics{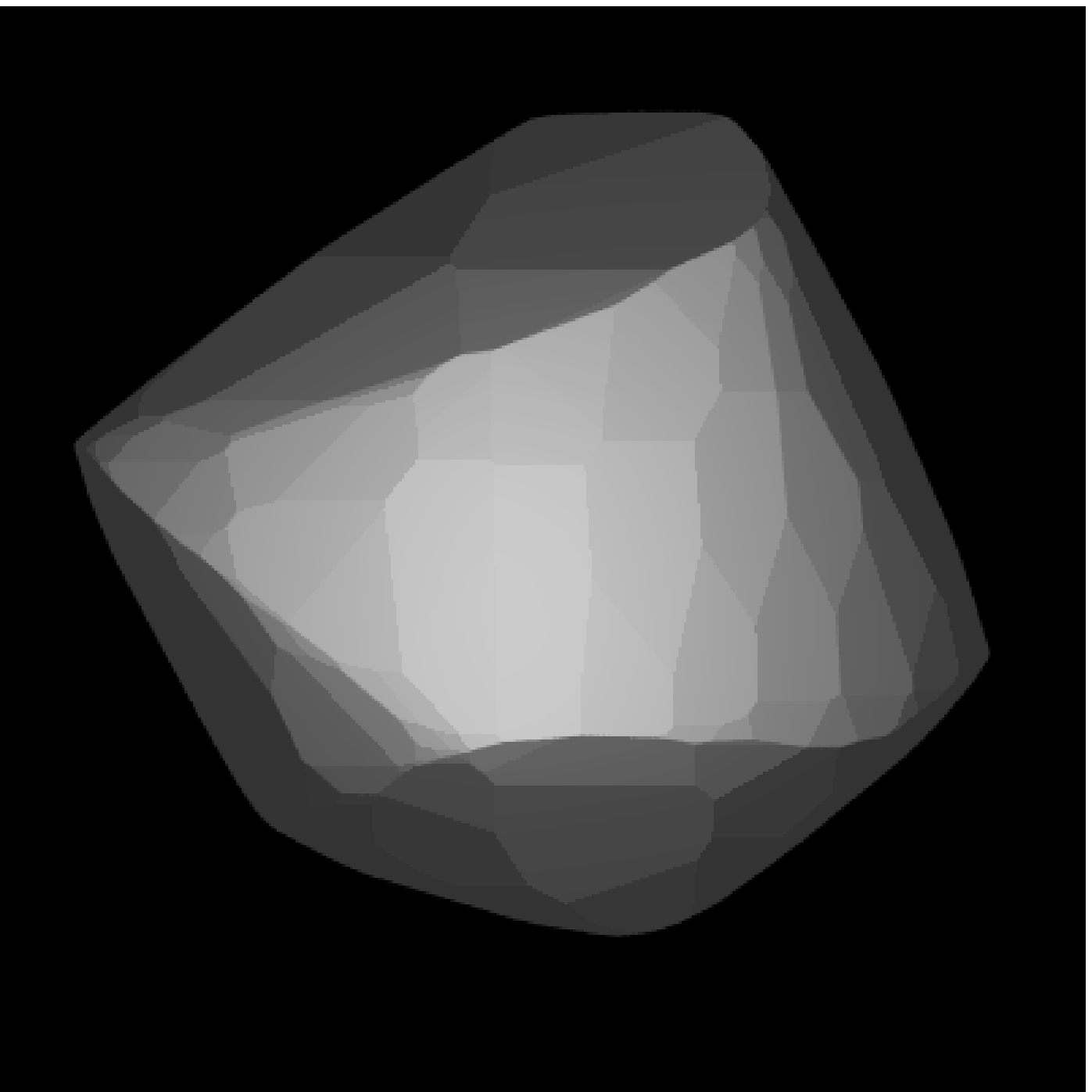}\includegraphics{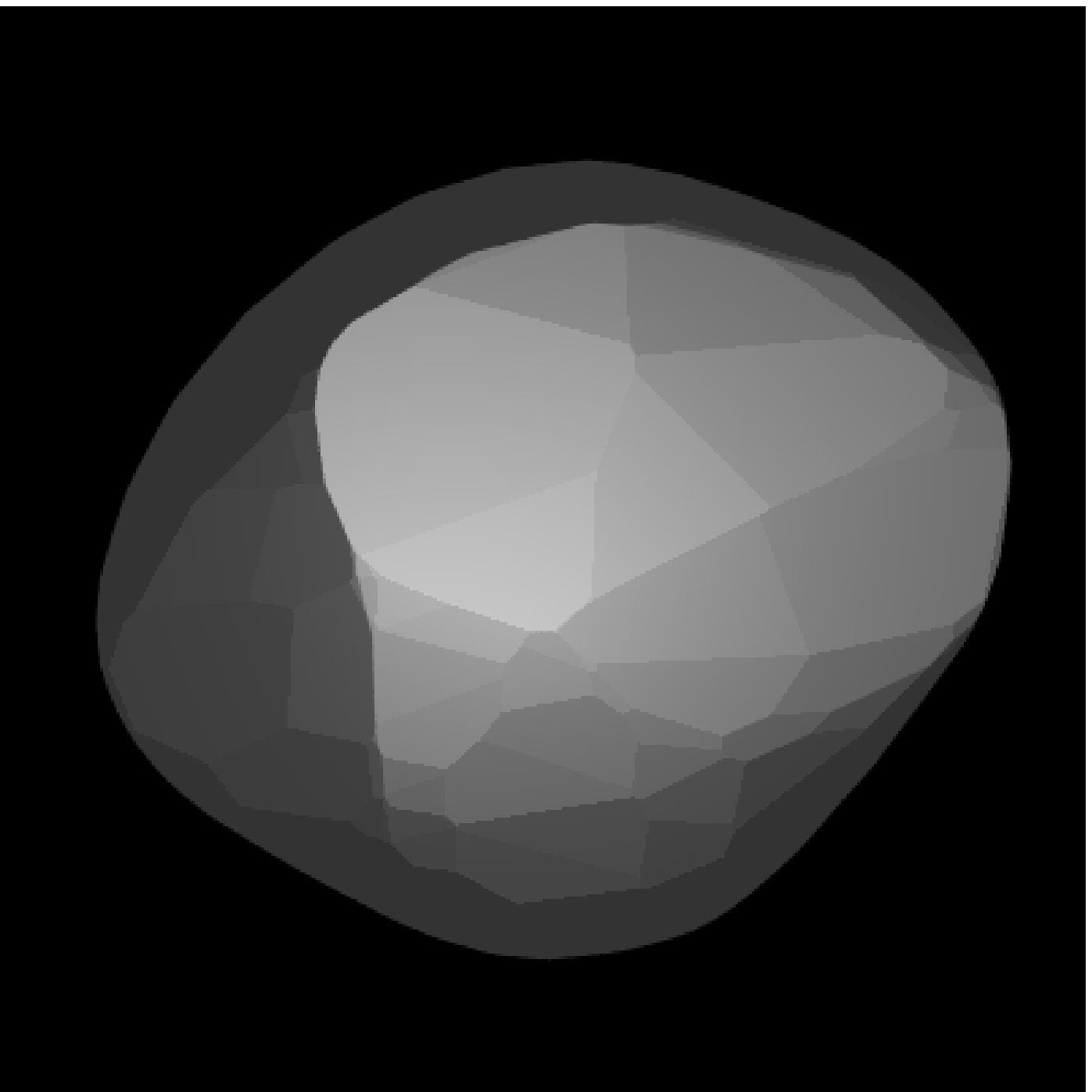}}\\
    \end{center}
    \caption{\label{fig:shape}Shape models that correspond to the first (top, $\lambda$=319$^{\circ}$) and second (bottom, $\lambda$=84$^{\circ}$) pole solutions derived from dense data alone. Each panel shows the shape model at three different viewing geometries: the first two are equator-on views rotated by 90$^{\circ}$, the third one is a pole-on view.}
\end{figure}

Optically dense data suggest a small light-curve amplitude of $\sim$0.1--0.15 magnitude, which is comparable with the typical noise of the available sparse-in-time measurements from astrometric surveys \citep{Hanus2011,Durech2016}. This implies that the sparse-in-time data should be dominated by noise. On the other hand, we have a large number of dense light-curves from many apparitions, which should be sufficient for a successful shape model determination, therefore we decided to use only those data. However, we excluded three light curves from our analysis because they had higher photometric noise (see Table~\ref{tab:photometry}). Moreover, we did not use the light curve from \citet{Wisniewski1997} for the final shape model determination because it disagreed with our best-fitting model: we obtained a significant offset of about 20 minutes in the rotation phase. Although we did not find any inconsistency in the original dataset of \citet{Wisniewski1997}, the error in the light curve cannot be fully ruled out. Other possible explanations involve, for example, effects of concavities or period change due to activity during 1989--1994 apparitions. However, our photometric dataset does not allow us to make reliable conclusions.

The fit in the rotation period subspace (we sampled periods in the proximity of the expected value that is well defined by the light-curve observations) exhibits one prominent minimum that corresponds to the period of $P=3.603958$~h (see Fig.~\ref{fig:per}). The corresponding $\chi^2$ is smaller by more than 10\% than those for all other periods, which is, according to our experience, sufficient to exclude all other solutions. Nevertheless, we show in Fig.~\ref{fig:periods} the comparison between fits of several light curves that correspond to the best and the second best periods. Next, we densely sampled various initial pole orientations: we ran the convex inversion with the best-fitting period and all the initial pole guesses. We obtained two pole solutions that fit the optical data significantly better than all the others. The first solution provides a slightly better fit than the second one, therefore the former is favored (the difference in $\chi^2$ is only about 5\%). Nevertheless, we still consider the second solution below because it is a plausible solution. The remaining pole solutions have $\chi^2$-values higher by more than 15\% and were therefore excluded. We show the comparison between fits of several light curves that correspond to the first, second, and third best poles in Fig.~\ref{fig:poles}. The resulting rotation state parameters are listed in Table~\ref{tab:param}, and the three-dimensional shapes at different viewing geometries are shown in Fig.~\ref{fig:shape}. The uncertainties in the pole direction, estimated by analyzing the dispersion of 30 solutions based on bootstrapped photometric data sets, are $\sim$5$^{\circ}$
for both solutions.

We have two poles with ecliptic latitude $\beta$ of about --39$^{\circ}$ and different ecliptic longitudes $\lambda$ (319$^{\circ}$, and 84$^{\circ}$). Moreover, their shapes seem to be different, therefore these solutions cannot be considered as mirror. The presence of two pole solutions is most likely caused by the high number of light curves with low amplitude, high noise, low period sampling, short time-span, and by an observational effect (not enough distinct viewing geometries). %However, as mentioned earlier, the first solution is slightly preferred.

The shape solution of \citet{Ansdell2014} ($P$=3.6032$\pm$0.0008 h, $\lambda$=85$\pm$13$^{\circ}$, $\beta$=--20$\pm$10$^{\circ}$, see also Table~\ref{tab:param}) is close to our second pole solution, but matches only when the extreme values of the parameter uncertainties
are considered. Our period estimate has a significantly higher precision because we found a global minimum in the period space. Ansdell's interval of periods includes multiple local minima (see their Fig.~3), which means that we cannot speak about a unique solution in their case. In principle, a best pole solution exists for each local minimum that does not necessarily need to be the same within different local minima. The correct approach would be to check all the possible pole solutions for each initial period (local minimum) and hope for only a few pole solutions to repeat. Ansdell et al. performed a statistical approach, where they determined the best-fitting period and associated error using a Monte Carlo technique. They added random Gaussian-distributed noise scaled to typical photometry errors of $\sim$0.01 mag to each light-curve point and modified the photometric data multiple times by generating each data point from a random distribution around its observed value taking the photometric uncertainties as dispersions. The main caveat is that this approach only takes the best-fitting period value, thus does not necessarily sample all local minima. As a natural consequence, some of the acceptable solutions could be easily missed. Ansdell et al. created a histogram of pole solutions that were derived based on different modified datasets and chose the most frequent solution. However, it is clear from their Figs.~4~and~5 that other pole solutions (e.g., those with $\lambda\sim$320$^{\circ}$) have rms comparable with their best solution. Our conclusion is that the photometric dataset used in \citet{Ansdell2014} was not sufficient for a unique shape model determination, and the reported pole is only one out of many possible solutions.

Our spin solution does not match the one of \citet{Krugly2002}, mainly because they reported a value of $\sim$--10$^{\circ}$ for the ecliptic latitude. Nevertheless, their rather preliminary pole solutions based on an ellipsoidal shape model assumption are relatively close to both our pole solutions and to the one of \citet{Ansdell2014} (Table~\ref{tab:param}).

The overall shape that corresponds to the first pole solution could evoke similarity to the spinning top shapes of some near-Earth primaries \citep[e.g., asteroid (66391) 1999 KW$_4$,][]{Ostro2006}, however, the equatorial ridge in our case is not that obvious and symmetric. On the other hand, the second shape model is more angular and seems to us rather less realistic.

\subsection{Thermophysical properties and size}\label{sec:TPMresults}

\begin{figure*}
    \begin{center}
        \resizebox{1.0\hsize}{!}{\includegraphics{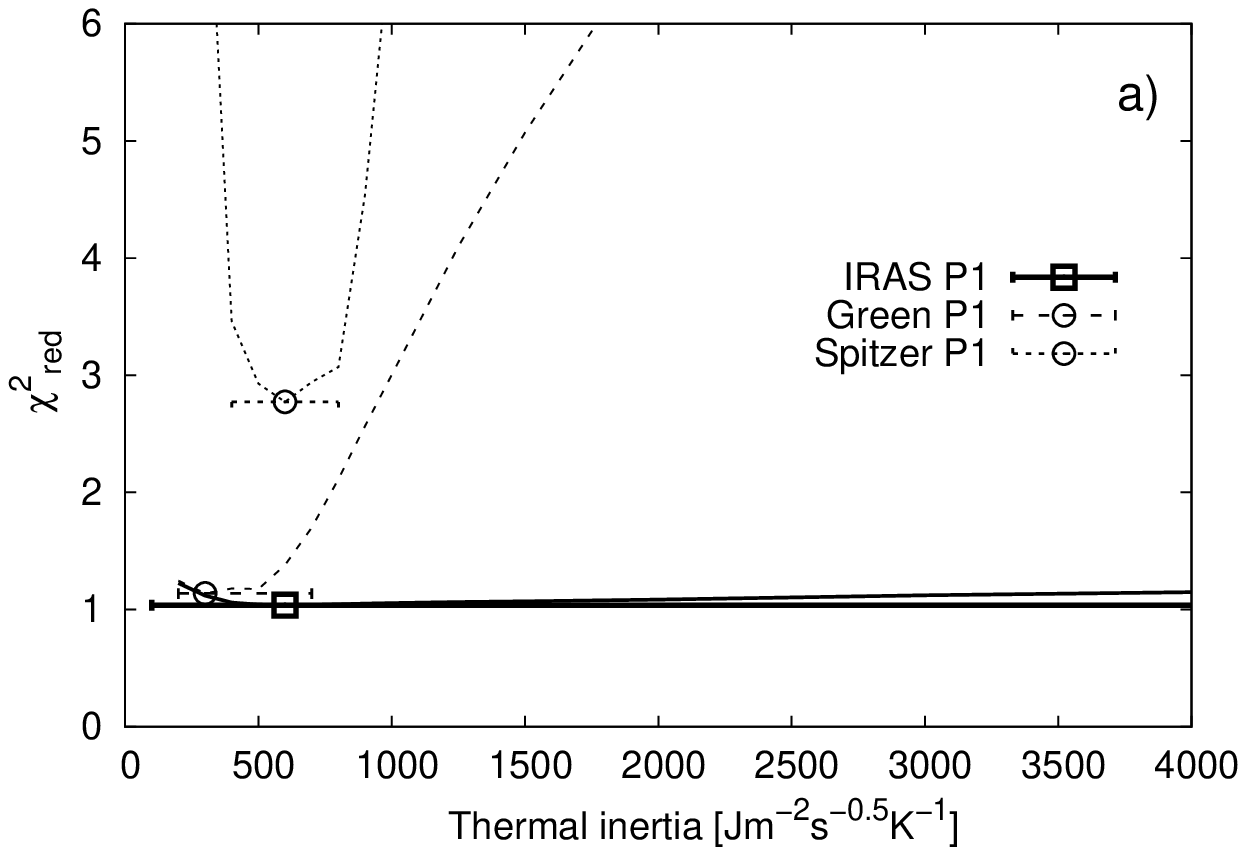}\includegraphics{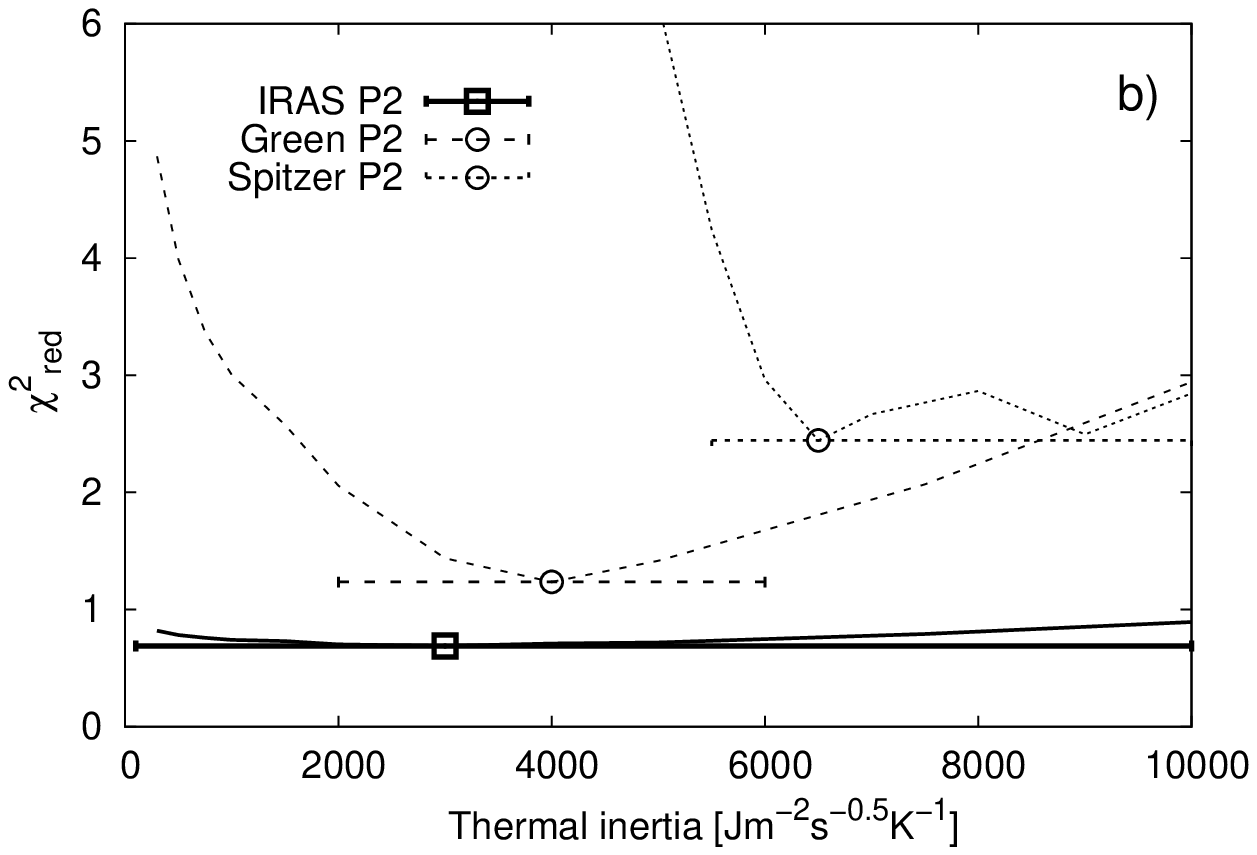}}\\
        \resizebox{1.0\hsize}{!}{\includegraphics{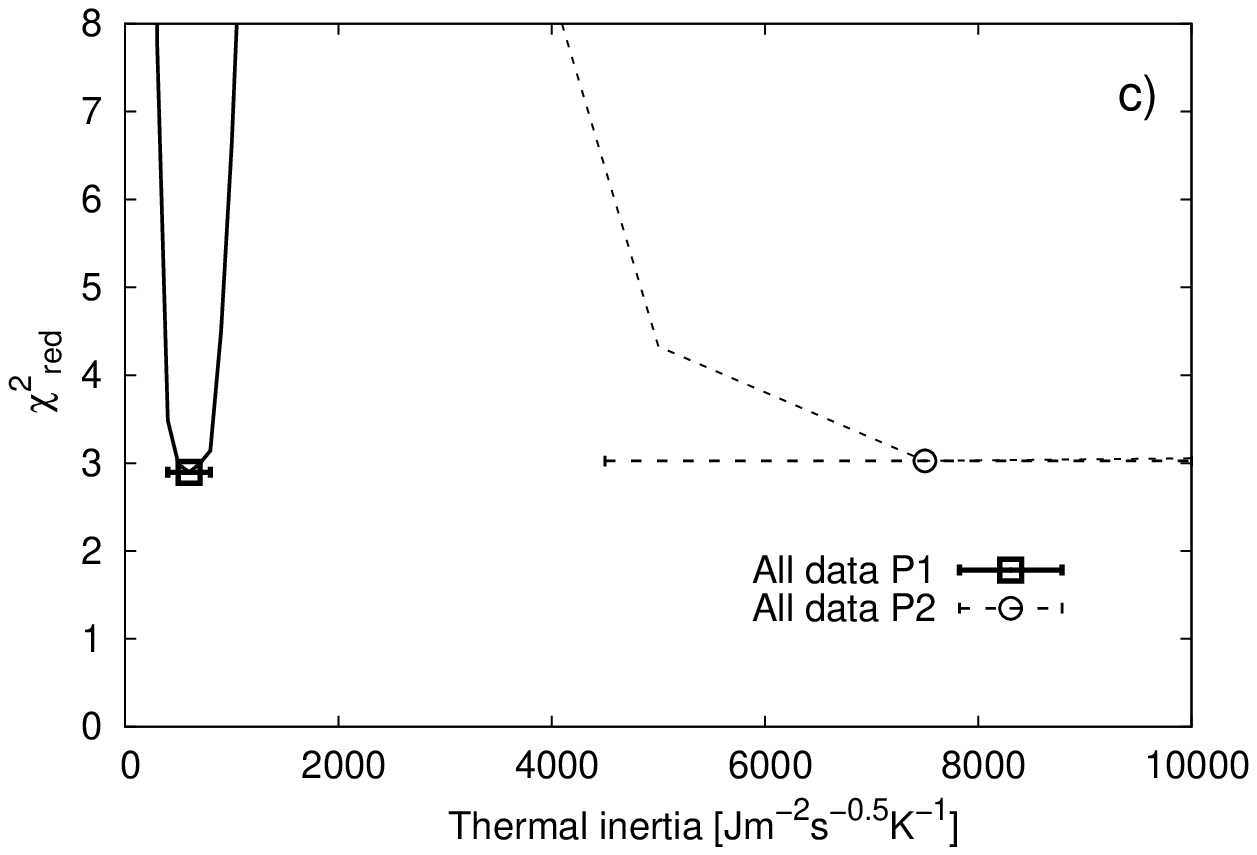}\includegraphics{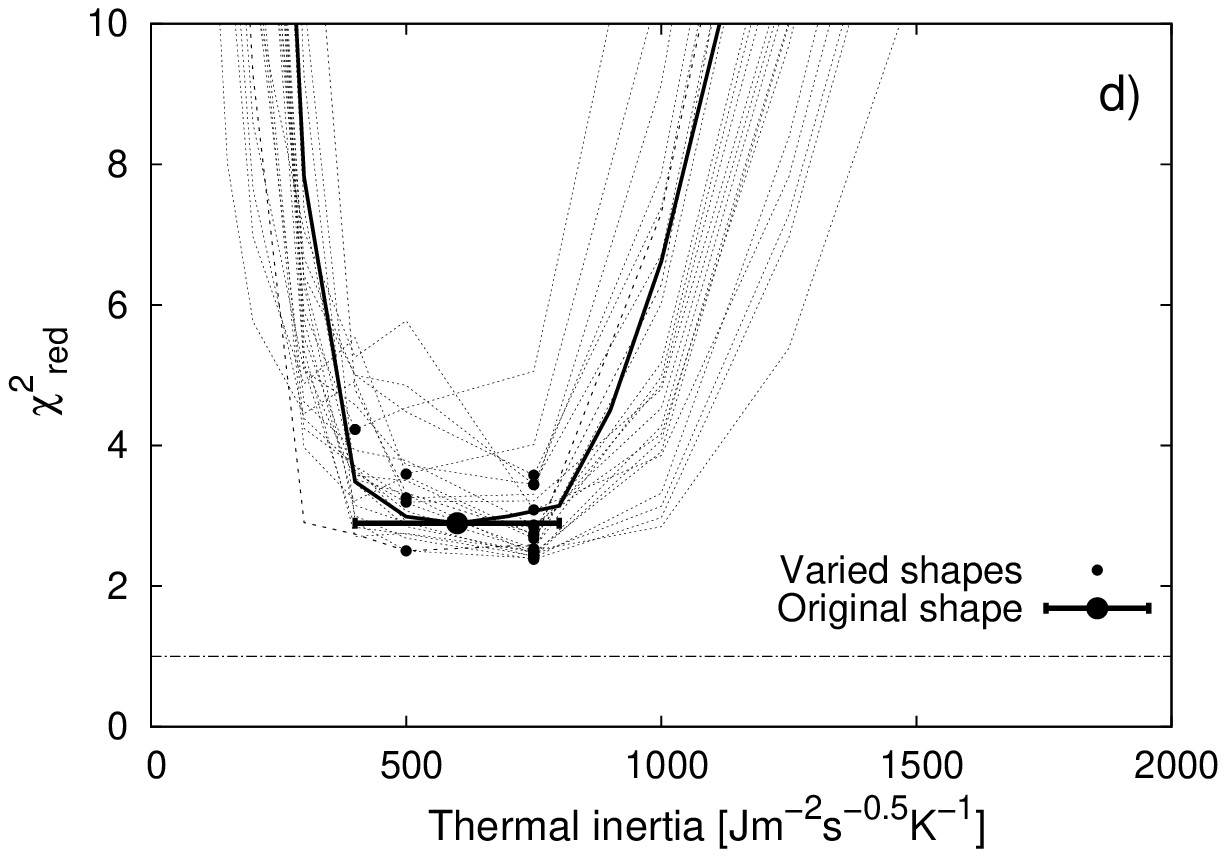}}\\
    \end{center}
    \caption{\label{fig:TPM}Thermophysical fit in the thermal inertia parameter subspace: a)~for all three thermal IR datasets (i.e., IRAS, Green, and Spitzer) individually with the first shape model as input, b)~for all three thermal datasets individually with the second shape model as input, c)~for the combined thermal dataset with both shape models as inputs, and finally d)~for the combined thermal IR dataset and the original and the nominal shape model with 29 close variants (we show only the results based on the first, strongly preferred, shape model).}
\end{figure*}

As described in Sect.~\ref{sec:IR}, we obtained three different datasets of thermal infrared measurements, namely from the IRAS satellite, the work of \citet{Green1985}, and the Spitzer space telescope. 

The absolute magnitude $H$ and slope parameter $G$ are necessary input parameters for the thermophysical modeling. $G$ is used to compute the geometric visible albedo $p_\mathrm{V}$ from the bolometric Bond albedo $A$ \citep[$A\approx (0.290+0.684G)\,p_{\mathrm{V}}$,][]{Bowell1989}, which is one of the fitted parameters in the TPM, and $H$ is a connection between $p_{\mathrm{V}}$ and size $D$. There are several often inconsistent values of $H$ and $G$ reported in the literature, thus our choice of reliable values needs careful justification. Values of $H$ from MPC (14.6), AstDyS (14.17) and JPL (14.51) are provided with an assumed value of $G$ (0.15). 
%
%The MPC value was used for the size determination from the IRAS and \citet{Green1985} data. 
%
However, these absolute magnitudes are based on astrometric data from sky surveys that have poor photometric accuracy. Additionally, \citet{Pravec2012a} found that absolute magnitudes (for asteroids with $H>14$) derived from astrometric surveys have an average systematic offset of about $-$0.4 magnitude.

Our light curves obtained during three nights in Ond\v rejov in 2004 (see Tab.~\ref{tab:photometry}) were calibrated in the Cousins R system and span phase angles of 12.2--28.0 deg. Based on these data, values of $H_R$=13.93$\pm$0.04 and $G$=0.15$\pm$0.03 were derived. To transform the magnitude to the V filter, we derived the Johnson-Cousins V--R color index from the visible spectra of \citet{Licandro2007} as 0.331. Moreover, the V--R color index of Phaethon was also derived by \citet{Skiff1996}, \citet{Dundon2005}, and \citet{Kasuga2008}: 0.34, 0.35$\pm$0.01, and 0.34$\pm$0.03, respectively. Applying a mean of all four values (0.34$\pm$0.01) resulted as $H$=14.27$\pm$0.04.

Another reliable absolute magnitude determination was reported by \citet{Wisniewski1997}, where the authors observed Phaethon in the Johnson V filter at a phase angle of 21.6 deg. They provided $H$=14.51$\pm$0.14 with an assumed $G$=0.23$\pm$0.12. We corrected their $H$ value to our $G$ parameter and obtained $H$=14.41$\pm$0.06.

Finally, we computed a weighted mean from these two estimates and used $H$=14.31 and $G$=0.15 in the TPM modeling.

To be complete, \citet{Ansdell2014} reported absolute magnitude and slope from their optical data obtained in the Cousins R filter. However, they claimed that they were unable to calibrate several epochs because Phaethon was not in a Sloan field, but those data seem to be included in the phase curve fit. We decided not to use these estimates of $H$ and $G$ since the calibration of the data could not be verified. Nevertheless, their value of $H_R$=13.90 is consistent with our determination of $H_R$=13.93$\pm$0.04.

As a first step, we decided to apply the TPM to all three thermal datasets individually. This should allow us to validate the quality of each dataset and potentially detect any inconsistency in the data.

IRAS fluxes have large uncertainties of $\sim$10--20\%. These values in absolute terms are even higher than the expected variability of the thermal light-curve because of the rotation. As a result, the thermophysical modeling did not reasonably constrain any of the desired parameters. The fit in the thermal inertia subspace is shown in panels a) and b) of Fig.~\ref{fig:TPM}. For both shape models, we obtained a TPM fit with a reduced $\chi^2$ of $\sim$1, where values of thermal inertia from $\sim$100 to several thousand \tiu are statistically indistinguishable. The size and albedo are constrained only poorly.

The flux uncertainties in the \citet{Green1985} dataset are usually $\sim$5--10\%, which proved to be sufficient to weakly constrain the thermophysical properties of Phaethon. The fit in thermal inertia for both shape models is shown in panels a) and b) of Fig.~\ref{fig:TPM}. Surprisingly, each shape model provides a different thermal inertia: the TPM fit with the first pole solution is consistent with values of $\Gamma\sim$200--700 \tiu, but the fit with the second one suggests $\Gamma>$2\,000 \tiu. Such high values are suspicious because they are significantly higher than measured for any other asteroid \citep{DelboAIV2015}. In addition, \citet{Opeil2010} provided upper limits for $\Gamma$ of CM and CK4 chondritic meteorite materials, both proposed as likely analogs of Phaethon. Thermal inertia values for either CM ($<$650 \tiu) and CK4 ($<$1\,400 \tiu) are much lower than our lower limit of $\sim$2\,000 \tiu derived for the second pole solution. Because the first shape model was already slightly preferred based on the light-curve dataset, the unusually high thermal inertia of the second model further supports this preference. We note that the derived geometric visible albedo of $\sim$0.14 (second pole solution) is higher than all previously reported values ($\sim$0.11), and that the best-fitting parameters correspond to a fit with a reduced $\chi^2$ of $\sim$1.

The quality of thermal data from Spitzer compared to those from IRAS and \citet{Green1985} is superior. As expected, the TPM fit therefore constrained thermal inertia, size, albedo, and surface roughness quite well. However, the minimum reduced $\chi^2$ is $\sim$3, but this can be attributed to weak absorption and emission features in the spectra that are not included in our TPM model because our TPM models the thermal IR continuum. The interpretation of the emission spectra is an object of our forthcoming publication. We note that the thermal inertia derived from the Spitzer data is generally similar to the one from the Green et al. data, but the physical properties are better constrained. The thermal inertia is $\sim$400--800 \tiu for the first pole solution and $>$3\,000 \tiu for the second one. Geometric visible albedo ($\sim$0.12) and size ($\sim$5.1~km) are consistent with previous estimates from the IRAS, AKARI, and WISE measurements. Medium values for the macroscopic surface roughness are preferred by the TPM (same as for the Green et al. data). Again, the first pole solution is preferred because it has more realistic thermal inertia. We note that the geometric visible albedo of the second pole solution is again rather high ($\sim$0.14).

All derived thermophysical parameters are listed in Table~\ref{tab:TPM}. The parameter uncertainties reflect the range of all solutions that have reduced $\chi^2$ values within the 1-sigma interval. Moreover, we also accounted for the uncertainty in the $H$ value (estimated as $\pm$0.05 mag) that contributes to the albedo uncertainty by $\pm$0.006. Additionally, systematic errors in the model (i.e., light-curve inversion, TPM) and data probably also affect parameter uncertainties. However, their contribution is difficult to estimate and thus is not accounted for in our final values. 

Owing to the superior quality of the Spitzer data, the TPM fit of combined Spitzer, IRAS, and Green et al. thermal data is similar to the fit with Spitzer data alone (see the fit in thermal inertia in panel c) of Fig.~\ref{fig:TPM} and parameter values in Table~\ref{tab:TPM}). 

Motivated by the findings of \citet{Hanus2015a} that the TPM results might be affected by the uncertainty in the shape model, we performed the varied-shape TPM modeling to check the stability of our TPM results based on fixed shape models as inputs. We bootstrapped the photometric dataset and constructed 29 shape models (by light-curve inversion) that map the uncertainty in the shape and rotation state. As we have shown that the second pole solution or shape model can be rejected (because it results in unreasonably high thermal inertia), we performed the TPM only with the 29 shapes that correspond to the first shape solution. The fits in thermal inertia for the 29 bootstrapped or varied shape models, as well as for the original one, are shown in panel d) of Fig.~\ref{fig:TPM}. The best-fit thermal inertias all cluster around the original value $\sim$600, indicating that this result is robust against shape uncertainties.  Nevertheless, including shape uncertainties does propagate through to slightly larger uncertainties in the fitted parameters.

To summarize, we derived a size of $D$=5.1$\pm$0.2~km that is consistent with previous estimates. Moreover, our geometric visible albedo of $p_{\mathrm{V}}$=0.12$\pm$0.01 is close to those reported ($\sim$0.11) as well. Our value of thermal inertia of $\Gamma$=600$\pm$200 J\,m$^{-2}$\,s$^{-1/2}$\,K$^{-1}$ agrees with those typical for small near-Earth asteroids with sizes ranging from a few hundred meters to a few kilometers, maybe slightly on the high end \citep{DelboAIV2015}. The intuitive interpretation of the relatively high thermal inertia value is that small bodies had short collisional lifetimes and accordingly developed only a coarse regolith \citep{Delbo2007a}. On the other hand, larger objects with much longer collisional lifetimes had enough time to build a layer of fine regolith, which results in lower values of thermal inertias. The close perihelion distance might lead to solar wind fluxes for Phaethon
that are high enough to directly remove any fine regolith from the surface. This could also explain the observed brightening near perihelion that is due to dust particles with an effective diameter of $\sim$1~$\mu$m \citep{Jewitt2013}.

\subsection{Orbital and spin axis evolution of Phaethon}\label{sec:dynamics}

Orbital and rotational motion of celestial bodies are generally not independent. Many examples of their mutual coupling have been discussed in planetary astronomy \citep[e.g., the spin vector alignments of asteroids in the Koronis family, or the spin state study in the Flora family,][]{Vokrouhlicky2003, Vrastil2015}. Here we deal with one particular aspect of these effects, namely how orbital motion influences long-term evolution of rotational motion. Our aim is to make use of our pole solutions from Sect.~\ref{sec:shapeModel} and investigate the conditions of the harsh surface irradiation near the pericenter of Phaethon's orbit during the past tens of kyrs.

\subsubsection{Orbital evolution}\label{sec:orbit}

\begin{figure*}
    \begin{center}
        \resizebox{0.5\hsize}{!}{\includegraphics{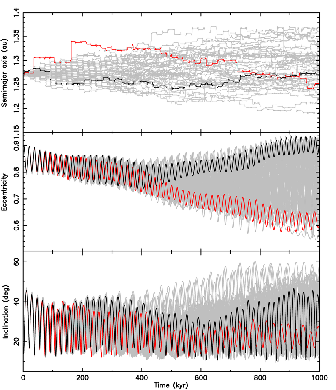}}\resizebox{0.5\hsize}{!}{\includegraphics{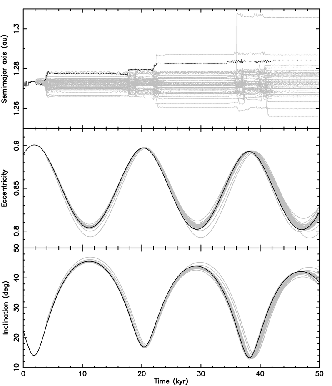}}\\
    \end{center}
    \caption{\label{fig:dynamics}Past dynamical evolution of Phaethon's orbit: (i) semimajor axis (top), (ii) eccentricity (middle), and (iii) inclination (bottom). Left panels show the whole integrated time-span of $1$~Myr, right panels show just the first $50$~kyr -- in both cases time goes to the past. We show the nominal orbit (black line), and also orbits of the $50$ clones (gray lines). The true past orbital evolution might have been any of those histories. Moreover, red lines represent clone evolution with an eccentricity that steadily decreases to the past.}
\end{figure*}

To start our analysis we first need to obtain information about Phaethon's orbit evolution. For the sake of our argument we are only interested in a certain time interval before the current epoch (somewhat arbitrarily, we chose $1$~Myr). We are aware of a strong chaoticity of the orbit evolution that is due to planetary encounters. Therefore, our results are just examples of possible past orbital histories of Phaethon. It is, however, important to note that the secular spin evolution is mostly sensitive to the orbital inclination and nodal longitude evolution, which are less strongly affected by the random component in planetary effects. 

We used the well-tested software package {\tt swift}\footnote{{\tt http://www.boulder.swri.edu/{\~{ }}hal/swift.html}.} to numerically integrate the nominal orbit of Phaethon, its clone variants, and the planets. All bodies were given their initial conditions at epoch MJD 57\,400: planets from the JPL DE405 ephemerides and the asteroid from the orbit solution provided by the {\tt AstDyS} website maintained at the University of Pisa. The asteroid close clones were created using the full covariant matrix of the {\tt AstDyS} solution. Because of Phaethon's rich observational record, the close clones differ from the nominal orbit by only very tiny values in all orbital elements (for instance, $\sim 2\times 10^{-9}$ in semimajor axis, $\sim 2\times 10^{-8}$ in eccentricity or $\sim 3\times 10^{-7}$ in inclination, all fractional values). As a result, the nominal solution does not have a special significance, and Phaethon might have followed any of the clone orbits. The backward integration in time was achieved by inverting the sign of velocity vectors. We used a short time-step of $\text{three}$~days and output state vectors of all bodies every $50$~years. This is a sufficient sampling for the spin vector integrations described in Sect.~4.3.2. We integrated all bodies to $1$~Myr. We included gravitational perturbations from planets, but neglected all effects of non-gravitational origin (such as recoil due to ejection of dust or gas particles and the Yarkovsky effect). For our purposes, however, this approximation is sufficient.

The evolution of Phaethon's orbit is shown in Fig.~\ref{fig:dynamics}. The longer timescale presented in the left column of plots shows that the orbital semimajor axis evolution is dominated by random-walk effects induced by the brief gravitational tugs that are caused
by planetary encounters. The evolution of eccentricity and inclination is different in nature from that of the semimajor axis. While also indicating a significant divergence of clone orbits, especially past the $\sim 100$~kyr time mark, the planetary encounters do not produce noticeable perturbations directly in eccentricity or inclination. Instead, their effect on $e$ and $i$ is indirect
because it is due to changes of secular frequencies reflecting the semimajor axis accumulated perturbation. Given the semimajor axis chaoticity, the eccentricities and inclination may also undertake diverse evolutions on a long term. For instance, the eccentricity might have stayed very high (as in the case of the nominal solution, black line), or steadily decreased to the past (selected clone shown by the red line). Overall, however, we note that the eccentricity of Phaethon's orbit has been increasing on average for all clone solutions during the past $\sim 300$~kyr. 

The right column of plots in Fig.~\ref{fig:dynamics} shows the
behavior of the orbital solution for Phaethon and its clones on a much shorter timescale, onely the past $50$~kyr. The semimajor axis evolution still shows clear  evidence of planetary encounters with a significant onset of divergence some $4$~kyr ago. Eccentricities and inclinations, however, are much more stable. We may appreciate the indirect planetary effect discussed above. For instance, the principal secular frequency of the inclination and node is among the fastest for the nominal solution and slower for many of the clone solutions (the inclination solution for clones trails in time behind the nominal inclination oscillations). This is because the nominal solution incidentally has the largest semimajor axis, while most of the clones were scattered to lower semimajor axis values. About $2$~kyr ago, Phaethon's orbital eccentricity was highest, causing its perihelion value to reach $q\sim 0.126$~au, which is significantly lower than its current value $q\sim 0.14$~au. This has been noted before \citep[e.g.,][]{Williams1993}. Several authors studied the possibility of a recent formation of the Geminids stream, including that near 0~AD when the perihelion had its last minimum (see, e.g., \citealt{Ryabova2007} or, for a more detailed overview, \citealt{Jenniskens2006}). However, a detailed match of the observed activity of Geminids over years may require particle feeding over an extended interval of time \citep{Jewitt2015}. We also note that the perihelion distance might not be the only parameter relevant to Phaethon's activity. It is possible that the spin axis orientation, studied in Sect.~\ref{sec:spin}, plays an equally important role. The highest eccentricity values during the previous cycles (i.e., about $20$~kyr ago and $38$~kyr ago) were lower and the perihelion was comparable to its current value. As mentioned above, this trend of decreasing eccentricity continues to about $300$~kyr ago.

It is also interesting to consider asteroids (155140) 2005~UD and (225416) 1999~YC. These two objects have been discussed as Phaethon's twins in the literature \citep[e.g.,][]{Ohtsuka2006,Ohtsuka2008,Ohtsuka2009}, perhaps fragments chopped off a common parent body of Phaethon and the Geminid complex. Considering a more tightly related orbit of 2005~UD, we repeated our backward integration of a nominal orbit and $50$ close clones. We confirm the proximity of the orbital evolution with that of Phaethon. However, the difference in orbital secular angles (longitudes of node and perihelion) prevents separation of (155140) from Phaethon in the immediate past \citep[see also][]{Ohtsuka2006}. We estimate that these two objects might have separated from a common parent body $\sim 100$~kyr ago or, more likely, even before this epoch. This means
that these kilometer-sized bodies, possibly related to Phaethon, must have a genesis in a different event in history than the current Geminid stream. 

Additionally, \citet{Kinoshita2007} reported that 2005 UD has $(B - V), (V - R)$ and $(R - I)$ color indices similar to those of Phaethon. As they are very rare colors (bluish spectral slope), it appears unlikely that the two asteroids could be just a random coincidence; it instead supports the suggestion that they are genetically related. It is also interesting that the primary spin period (3.6 h) and
the size ratio (estimated from their absolute magnitude difference
of 3.0) of the asteroid pair Phaethon--2005 UD agrees with the model of the spin-up fission formation of the asteroid pair \citet{Pravec2010}. This suggests that 2005 UD was formed from material escaped from Phaethon after it was spun up to the critical rotation rate (presumably by the Yarkovsky-O'Keefe-Radzievskii-Paddack effect, YORP).

\subsubsection{Spin evolution}\label{sec:spin}

\begin{figure*}
    \begin{center}
        \resizebox{1.0\hsize}{!}{\includegraphics{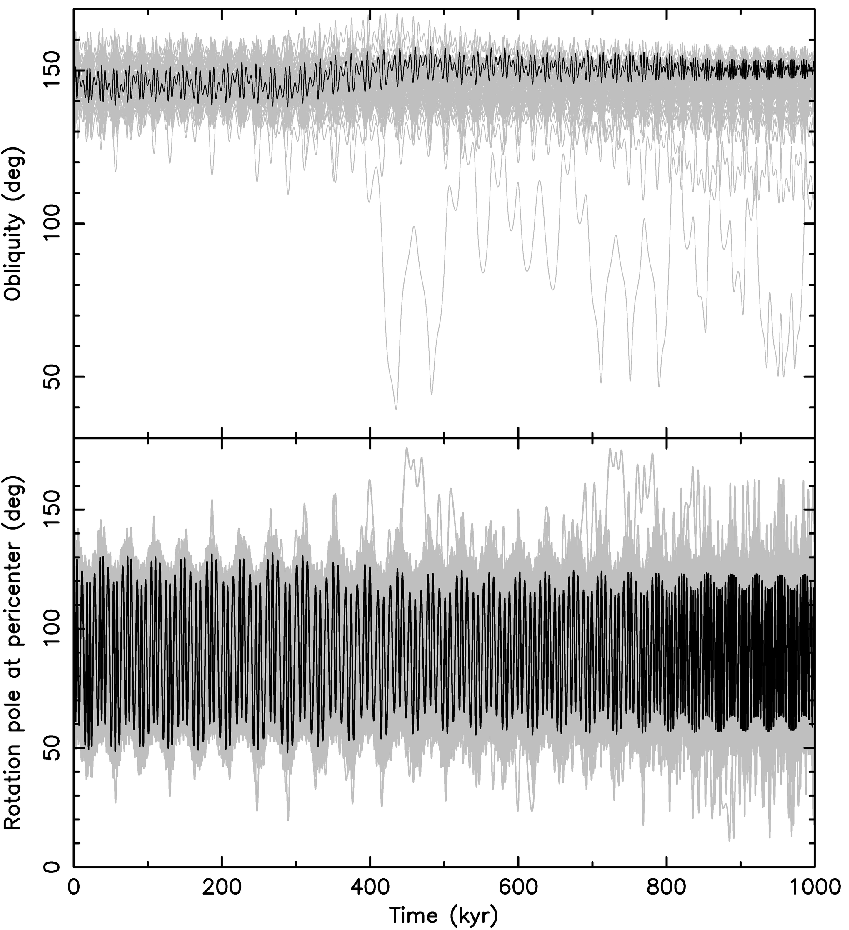}\includegraphics{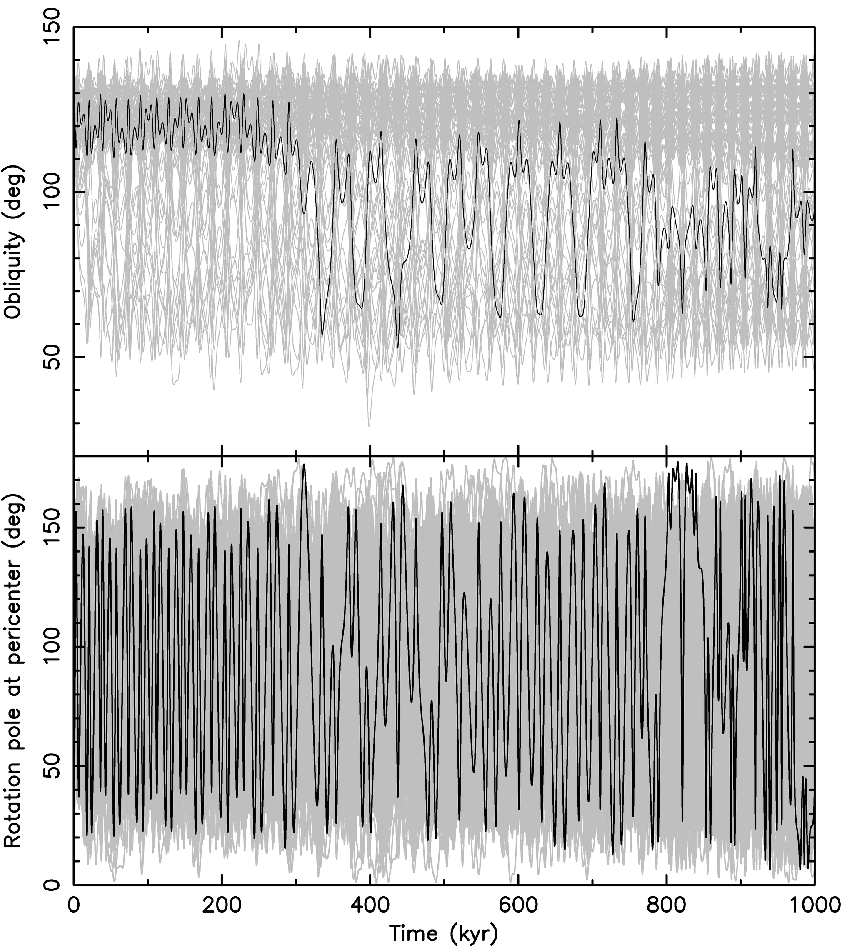}}\\
        \resizebox{1.0\hsize}{!}{\includegraphics{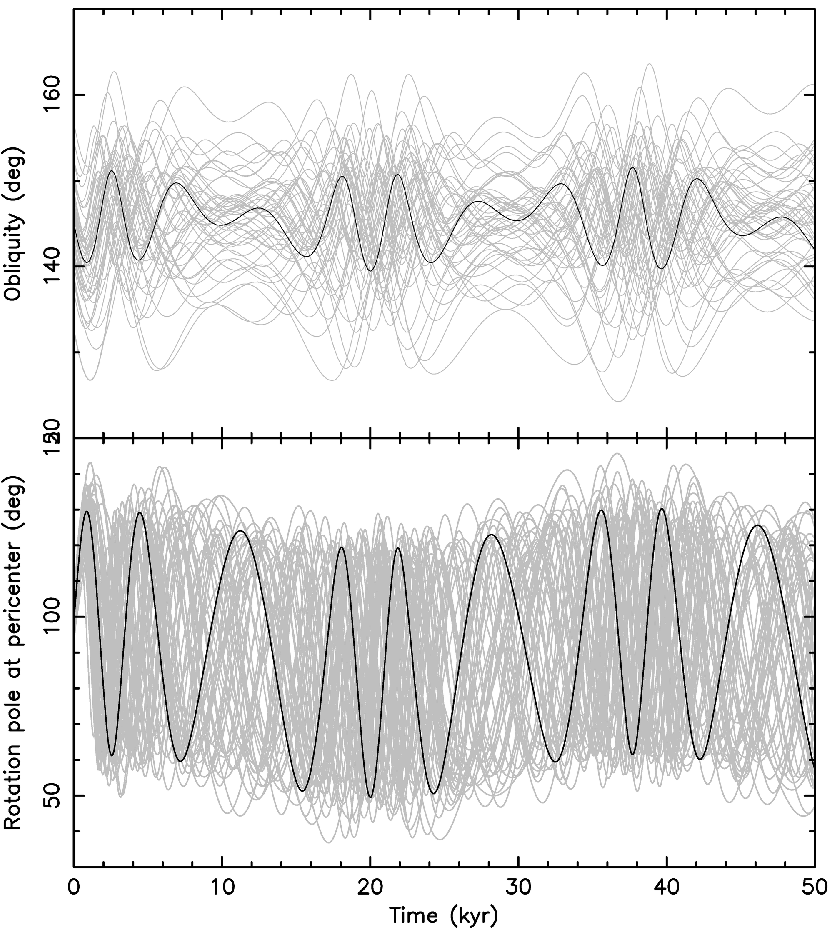}\includegraphics{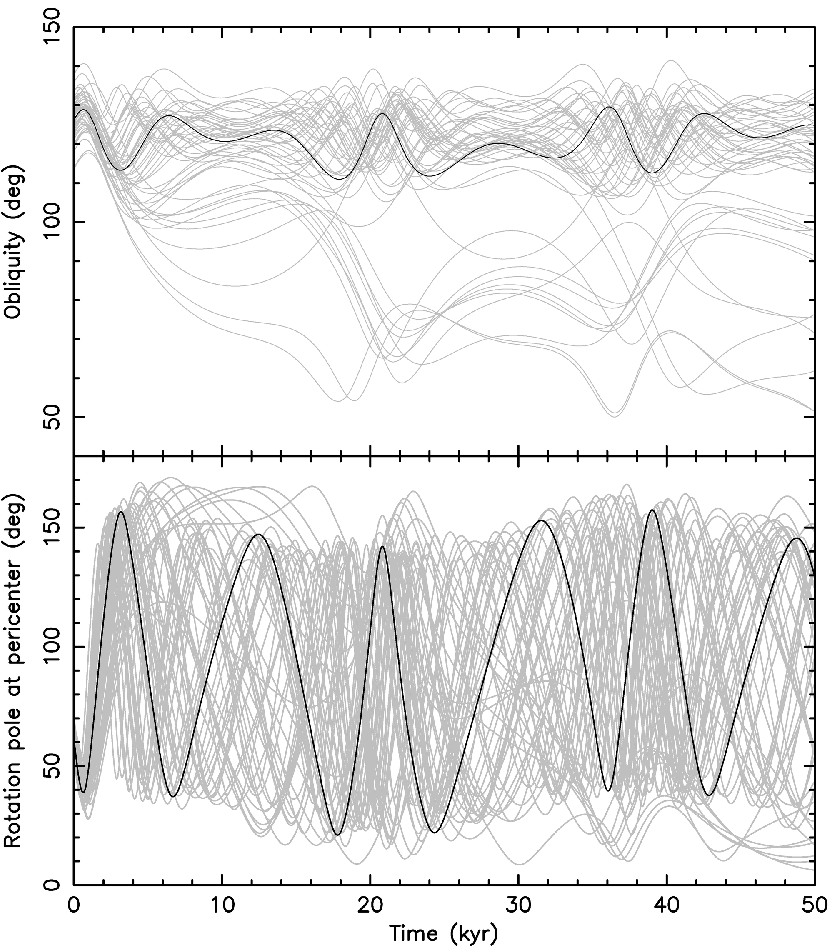}}\\
    \end{center}
    \caption{\label{fig:spin}Past dynamical evolution of Phaethon's spin axis direction ${\bf s}$: left column of panels for the P1 solution $(\lambda,\beta)=(319^\circ, -39^\circ)$, right column of panels for the P2 solution $(\lambda,\beta)=(84^\circ, -39^\circ)$. The upper two panels show the spin evolution over the whole integrated interval of $1$~Myr, the bottom two panels zoom at the past $50$~kyr. Each time the nominal solution is shown by the black line, while the gray lines are for the $50$ spin clones (see the text). Each of the four sections shows (i) the obliquity $\varepsilon$ at the top (angle between ${\bf s}$ and normal to the osculating orbital plane), and (ii) the angle $\alpha$ between ${\bf s}$ and the direction to the Sun at pericenter.}
\end{figure*}

We used the secular model of the asteroid rotation-state evolution formulated in \citet{Breiter2005}. In this framework, all dynamical effects with periods shorter than rotational and orbital periods are eliminated by averaging, and the spin evolution is considered on long timescales. This is not only sufficient for our work here, but it also considerably speeds up numerical simulations. We included the gravitational torque due to the Sun and, again, neglected all effects of non-gravitational origin. This implies that the mean rotation period is conserved and the evolution is principally described by changes in direction ${\bf s}$ of the spin axis. The importance of the model arises from the fact that the heliocentric orbit evolves by the planetary perturbations, as has been described in the previous section. The characteristic timescale of the spin axis precession in space that is due to the solar torque may be similar to that of the orbit precession about the pole of ecliptic. If so, interesting resonant phenomena may occur and produce a complicated evolution of the asteroid's rotational pole \citep[e.g.,][]{Vokrouhlicky2006}.

Our torque model is only approximate and needs to be made more accurate in the future if observations were to require it. In particular, the Yarkovsky-O'Keefe-Radzievskii-Paddack effect
(YORP) might be estimated when (i)~the shape model and spin state is improved, and (ii)~photometry over a longer timespan is available. At present, our conclusions and test runs do not require the
YORP effect to be included in the model. For instance, using our preferred pole and shape model for Phaethon, we estimate that in $1$~Myr the rotation period would change by a few percent and the obliquity by only less than ten degrees. We used the simple one-dimensional model presented in \citet{Capek2004} with
a bulk density of $1$ g~cm$^{-3}$. The main factor that weakens the YORP effect is Phaethon's large size.

Conveniently, \citet{Breiter2005} also provided an efficient symplectic integration scheme for the spin axis evolution. In
addition to the initial direction of ${\bf s}$ at some chosen epoch, the model needs two ingredients. First, the orbit evolution that is due to planetary perturbations is required. In our case, this is provided by the numerical integrations above. Second, we need to know the precession constant of the asteroid. For a given rotation period its value depends on a single parameter, usually called dynamical ellipticity $\Delta = (C - 0.5 (A+B))/C$, where $A$, $B$, and $C$ are principal moments of the inertia tensor of the body. In principle, $\Delta$ can be estimated from our shape models obtained by the light-curve inversion in Sect.~\ref{sec:shapeModel}. Our nominal models for both pole solutions yield $\Delta\simeq 0.11$. However, shape variants that are still compatible with the fit to available light curves could  provide $\Delta$ values in the $0.06$ to $0.16$ interval. We used this range as the uncertainty of Phaethon's $\Delta$ value.

Propagation of Phaethon's spin vector ${\bf s}$ is subject to uncertainty that originates from several sources. First, the initial conditions are not exact. We assumed a $5^\circ$ uncertainty in both ecliptic longitude and latitude of our two nominal pole solutions from Sect.~\ref{sec:shapeModel}. Because we found that one of the nominal pole solutions provides a more satisfactory thermal inertia value (thereafter denoted P1), we studied the spin histories starting in this solution's vicinity first. Then we proceed with those starting in the vicinity of the second nominal pole solution (thereafter denoted P2). Second, the dynamical ellipticity has $\simeq \pm 0.05$ uncertainty. Third, the orbit may follow any of the clone variants described in Sect.~\ref{sec:orbit}. Our tests show that the third issue is the least important (at least on the timescale we are interested in). For simplicity, we therefore assumed our nominal solution of Phaethon's orbital evolution alone and considered \textup{{\em \textup{spin clones}} }by propagating ${\bf s}$ from (i) different initial conditions, and (ii) with different $\Delta$ values. As in the case of orbital clones above, the past spin evolution of Phaethon may follow any of the clone solutions with equal statistical likelihood. We integrated the spin evolution to $1$~Myr backward in time, the same interval for which we integrated the orbit. The time
step was $50$~yr, sufficient to describe secular effects.

To communicate the results of the integrations in a simple way, we used two angular variables related to the ${\bf s}$ direction. First, we used the obliquity $\varepsilon$, namely the angle between ${\bf s}$ and the normal to the osculating orbital plane. The obliquity directly informs us about the sense of Phaethon's rotation, and this is important for some orbital accelerations of non-gravitational origin (such as the Yarkovsky effect). Second, we also determined the angle $\alpha$ between ${\bf s}$ and direction to the Sun \textup{\textup{\textup{{\em \textup{at perihelion}}} }}of the orbital motion. The angle $\alpha$ is an important parameter determining which hemisphere of Phaethon is preferentially irradiated near perihelion (e.g., $\alpha\ll 90^\circ$ implies that the
northern hemisphere receives most of the solar radiation at pericenter an vice versa).

Figure~\ref{fig:spin} shows our results. We focus first on our preferred pole solution P1 with an initial ecliptic longitude $\lambda\simeq 319^\circ$ (left column). The obliquity exhibits stable oscillations about the mean value $\sim 148^\circ$ with typically a small amplitude of $\sim 10^\circ$, much lower than the orbit inclination. This is because the spin axis precession frequency due to the solar gravitational torque is much faster than the orbit-plane precession frequency. Additionally, the sense of precession is opposite for retrograde rotation ($\varepsilon > 90^\circ$). Obliquity of just one exceptional spin clone shows irregular excursions to $\sim 40^\circ$, being thus temporarily prograde before returning to the retrograde-rotation zone. This phenomenon is due to the existence of a large chaotic zone associated with the Cassini secular resonance between the spin axis precession and orbital precession with a mean frequency of $\simeq -32.5$~arcsec/yr. While the nominal resonant obliquity is $\simeq 84.5^\circ$, the resonant zone extends from $\simeq 30^\circ$ to $\simeq 120^\circ$ obliquity \citep[see][for more examples]{Vokrouhlicky2006}. Therefore, the nominal pole solution P1 is barely safe from these chaotic effects. The angle $\alpha$ between ${\bf s}$ and the direction to the Sun at perihelion also shows rather stable oscillations about a mean value of $\simeq 90^\circ$. The past $50$~kyr of evolution (bottom left plots at Fig.~\ref{fig:spin}) show that the current value of $\simeq 98.8^\circ$ switches to about $60^\circ$ some $\simeq 2$~kyr ago. Therefore, when the perihelion was lowest ($\sim 0.126$~au), the northern hemisphere of Phaethon was harshly irradiated by sunlight at perihelion. But the values of $\alpha$ oscillate from $\simeq 50^\circ$ to about $\simeq 130^\circ$. This means that $\simeq 500$~yr ago, and again $\simeq 4$~kyr ago, it was the southern hemisphere's turn to be irradiated at perihelion. This probably is the reason why Phaethon's surface does not show any convincing hemispheric spectral or color asymmetry \citep{Ohtsuka2009}. Additionally, we wonder whether these polar-irradiation cycles, together with variations in perihelion distance, play a role in the strength of Phaethon's activity and thus possibly in the origin of different components of the Geminids stream.

Our second nominal pole solution with the initial ecliptic longitude $\lambda\simeq 84^\circ$ (right column in Fig.~\ref{fig:spin}) provides a wilder scenario. This is because the initial obliquity $\simeq 126^\circ$ is dangerously close to the chaotic layer of the above-mentioned Cassini resonance. Even the nominal spin solution shown by the black line is dragged to the prograde-rotation zone of obliquities $< 90^\circ$ at $\sim 300$~kyr ago. Many spin clones follow a similar evolution, although others stay safely in the retrograde-rotating region. As expected, the angle $\alpha$ between ${\bf s}$ and the solar direction at pericenter also exhibits much stronger oscillations than in the P1 solution. For P2 the geometry is switched, such that currently the southern hemisphere is preferentially irradiated at perihelion \citep[e.g.,][]{Ohtsuka2009,Ansdell2014}. However, $\simeq 2$~kyr ago the geometry changes for the P2 solutions and the northern hemisphere was irradiated at the perihelion.

Finally, we would like to set right a slight misconception from \citet{Ansdell2014}. The current $\varepsilon > 90^\circ$ obliquity of Phaethon, holding for either of the P1 and P2 solutions, cannot be directly linked to the Yarkovsky transport from the main belt. At the moment Phaethon's precursor asteroid left the main belt zone, as envisaged for instance by \citet{deLeon2010}, the rotation state was likely retrograde (only retrograde members of the Pallas collisional family can enter the 8:3 mean motion resonance, which corresponds to the most probable dynamical pathway of Phaethon). However, in millions of years when the orbit was evolving in the planet-crossing zone, the obliquity might have undergone chaotic evolution driven by Cassini spin-orbit resonances and the YORP effect. We note, for instance, that the nominal P2 solution in Fig.~\ref{fig:spin} indeed transits from a prograde to retrograde regime in the last $1$~Myr of evolution.

\section{Conclusions}\label{sec:conclusions}

Twenty light curves were presented here using six different telescopes (65cm in Ond\v rejov, UH88 in Hawaii, 60cm in Modra, 1m in Calern, IAC-80 at Teide, and 66cm at Badlands Observatory)  between November 2, 1994 and October 8, 2015.

We derived a unique shape model of NEA (3200)~Phaethon based on previous and newly obtained light curves. This model will be useful for planning the observations during the December 2017 close approach (as close as 0.069 au to Earth). Although two pole solutions are consistent with the optical data, only the formally better solution -- sidereal rotation period of 3.603958(2) h and ecliptic coordinates of the preferred pole orientation of (319$\pm$5, $-$39$\pm$5)$^{\circ}$ -- provides a TPM fit of the thermal (mid-)infrared data with realistic thermophysical parameters.  

Our newly obtained light curves, their comparison with the modeled light curves, and the shape model are available in the Database of Asteroid Models from Inversion Techniques.

By applying a thermophysical model to thermal fluxes from the IRAS satellite and \citet{Green1985}, and mid-infrared spectra from the Spitzer space telescope, we derived a size ($D$=5.1$\pm$0.2~km), a geometric visible albedo ($p_{\mathrm{V}}$=0.122$\pm$0.008), a thermal inertia ($\Gamma$=600$\pm$200 \tiu) and a medium surface roughness for Phaethon. These values are consistent with previous estimates. The derived thermal inertia is slightly higher than for those of similarly sized near-Earth asteroids, but they are still consistent.

Based on our study of a long-term orbital evolution of Phaethon, we confirm previous findings that about $2$~kyr ago Phaethon's orbital eccentricity was at its highest, causing its perihelion value to reach $q\sim 0.126$~au, which is significantly lower than its current value $q\sim 0.14$~au. The eccentricity of Phaethon's orbit has been increasing on average for all clone solutions during the past $\sim 300$~kyr. Both behaviors could be relevant for the origin of the Geminids. We note that the highest eccentricity values during the previous (oscillating) cycles (i.e., about $20$~kyr ago and $38$~kyr ago) were lower and the perihelion was comparable to its current value.

We confirm the proximity of the orbital evolution of asteroids (155140) 2005~UD and (225416) 1999~YC with that of Phaethon. However, the difference in orbital secular angles (longitudes of node and perihelion) precludes their separation from Phaethon in the immediate past \citep[see also][]{Ohtsuka2006}, but rather $\sim 100$~kyr ago or, more likely, even before this epoch. This
means that these kilometer-sized bodies are probably not related to the current Geminid stream. The similarities in color indices of Phaethon and 2005~UD and their size ratio further support the existence of the Phaethon--2005~UD asteroid pair.

The obliquity of the preferred spin solution exhibits stable oscillations about the mean value $\sim 148^\circ$ with typically a small amplitude of $\sim 10^\circ$, much lower than the orbit inclination. The existence of a large chaotic zone associated with Cassini secular resonance between the spin axis precession and orbital precession can temporarily switch the rotation to a prograde one. While this behavior is rare for the preferred pole solution, it is very common for the second pole solution.

When the perihelion was the lowest about 2~kyr ago ($\sim 0.126$~au), the northern hemisphere of Phaethon (preferred pole solution) was harshly irradiated by sunlight at perihelion. On the other hand, $\simeq 500$~yr ago, and again $\simeq 4$~kyr ago, it was the southern hemisphere's turn to be irradiated at perihelion. This probably explains the lack of any convincing hemispheric spectral or color variations \citep{Ohtsuka2009}.

\begin{acknowledgements}
JH greatly appreciates the CNES post-doctoral fellowship program. The work of DV and PP was supported by the Czech Science Foundation (grants GA13-01308S and P209-12-0229). The work of VAL is performed in the context of the NEOShield-2 project, which has received funding from the European Union's Horizon 2020 research and innovation programme under grant agreement No. 640351. MD and VAL were also supported by the project under the contract 11-BS56-008 (SHOCKS) of the French Agence National de la Recherche (ANR). The work at Modra is supported by the Slovak Grant Agency for Science VEGA (Grant 1/0670/13).

This article is based on observations made with the IAC-80 operated on the island of Tenerife by the Instituto de Astrof\'{\i}sica de Canarias in the Spanish Observatorio del Teide. JL acknowledges support from the project ESP2013-47816-C4-2-P (MINECO, Spanish Ministry of Economy and Competitiveness).

The authors wish to acknowledge and honor the indigenous Hawaiian community because data used in this work were obtained from the summit of Maunakea. We honor and respect the reverence and significance the summit has always had in native Hawaiian culture. We are fortunate and grateful to have had the opportunity to conduct observations from this most sacred mountain.

This work is based in part on observations made with the Spitzer Space Telescope, which is operated by the Jet Propulsion Laboratory, California Institute of Technology under a contract with NASA. Support for this work was provided by NASA through an award issued by JPL/Caltech.
\end{acknowledgements}

%\include{tab1}
%\include{tab2}
%\include{tab3}
%\include{tab4}

%\scriptsize{
\begin{table*}
\caption{\label{tab:photometry}List of dense-in-time light curves used for the shape modeling. For each light curve, the table gives the epoch, number of points $N_p$, asteroid distances to the Sun $r$ and Earth $\Delta$, used filter and telescope, observer, and reference.}
\centering
\begin{tabular}{ rlr rr r rrr}
\hline 
\multicolumn{1}{c} {N} & \multicolumn{1}{c} {Epoch} & \multicolumn{1}{c} {$N_p$} & \multicolumn{1}{c} {$r$} & \multicolumn{1}{c} {$\Delta$} & \multicolumn{1}{c} {Filter} & Telescope & Observer  & Reference \\
 &  &  & [au] & [au] &  &  &  &  \\
\hline\hline
  1 & 1989-10-09.4 & 30 & 2.02 & 1.22 & V & 90 inch Steward Observatory & Wisniewski & \citet{Wisniewski1997}, rejected \\
  2 & 1994-11-02.1 &  22 & 1.82 & 1.04 & R & D65 & Pravec &     This work    \\
  3 & 1994-12-02.9 &  14 & 1.53 & 0.56 & R & D65 & Pravec &     This work \\
  4 & 1994-12-04.1 &  17 & 1.51 & 0.54 & R & D65 & Pravec &     This work \\
  5 & 1994-12-06.9 &  13 & 1.48 & 0.52 & R & D65 & Pravec &     This work \\
  6 & 1994-12-27.3 &  76 & 1.22 & 0.44 & R & Lowell & Buie & \citet{Ansdell2014} \\
  7 & 1995-01-04.4 &  11 & 1.10 & 0.46 & R & UH88 & Meech, Hainaut & \citet{Ansdell2014} \\
  8 & 1995-01-04.8 &  45 & 1.10 & 0.46 & R & D65 & Pravec &   \citet{Pravec1998} \\
  9 & 1995-01-05.4 &  79 & 1.09 & 0.46 & R & UH88 & Meech, Hainaut & \citet{Ansdell2014} \\
 10 & 1997-11-01.1 &  88 & 1.32 & 0.78 & R & D65 & Pravec &   \citet{Pravec1998} \\
 11 & 1997-11-02.1 &  80 & 1.31 & 0.76 & R & D65 & Pravec &   \citet{Pravec1998} \\
 12 & 1997-11-11.6 &  39 & 1.18 & 0.56 & R & UH88 & Meech, Bauer  & \citet{Ansdell2014} \\
 13 & 1997-11-12.6 &  52 & 1.16 & 0.54 & R & UH88 & Meech, Bauer & \citet{Ansdell2014} \\
 14 & 1997-11-21.6 &  48 & 1.02 & 0.39 & R & UH88 & Meech, Bauer & \citet{Ansdell2014} \\
 15 & 1997-11-22.6 &  47 & 1.01 & 0.37 & R & UH88 & Meech, Bauer & \citet{Ansdell2014} \\
 16 & 1997-11-25.6 &  24 & 0.95 & 0.34 & R & UH88 & Meech, Bauer & \citet{Ansdell2014} \\
 17 & 1998-11-22.1 &  14 & 2.31 & 1.36 & R & IAC-80 & Licandro &  This work \\
 18 & 1998-11-23.1 &  16 & 2.31 & 1.36 & R & IAC-80 & Licandro &  This work \\
 19 & 1998-12-08.0 &   9 & 2.26 & 1.39 & R & IAC-80 & Licandro &   This work \\
 20 & 1998-12-09.0 &  15 & 2.25 & 1.40 & R & IAC-80 & Licandro &   This work \\
 21 & 2003-11-20.8 &  18 & 1.76 & 0.81 & R & D65 & Pravec, Ku\v snir\' ak & This work, rejected \\
 22 & 2004-11-13.3 &  12 & 1.84 & 0.89 & R & Badlands Observatory & Reddy, Dyvig &  This work, rejected  \\
 23 & 2004-11-19.5 &  38 & 1.78 & 0.83 & R & UH88 & Dundon &      \citet{Ansdell2014} \\
 24 & 2004-11-21.6 &  51 & 1.76 & 0.81 & R & UH88 & Dundon &      \citet{Ansdell2014} \\
 25 & 2004-11-22.4 &  35 & 1.75 & 0.80 & R & UH88 & Dundon &      \citet{Ansdell2014} \\
 26 & 2004-11-25.1 &  47 & 1.73 & 0.76 &   & Modra & Gal\'ad & This work, rejected \\
 27 & 2004-12-05.0 & 101 & 1.63 & 0.67 & R & D65 & Pravec, Ku\v snir\' ak &   This work  \\
 28 & 2004-12-05.3 &  41 & 1.63 & 0.67 & R & Badlands Observatory & Reddy, Dyvig &   This work  \\
 29 & 2004-12-11.0 & 148 & 1.57 & 0.64 & R & D65 & Pravec, Ku\v snir\' ak &    This work \\
 30 & 2004-12-18.8 &  15 & 1.48 & 0.61 & R & D65 & Pravec, Ku\v snir\' ak &    This work \\
 31 & 2007-11-17.2 &  47 & 1.28 & 0.51 &   & Modra & Gal\'ad & This work \\
 32 & 2007-11-28.2 &  96 & 1.13 & 0.29 &   & Modra & Gal\'ad & This work \\
 33 & 2007-12-04.1 & 232 & 1.03 & 0.18 &   & Modra & Korno\v s, Vil\' agi & This work \\
 34 & 2013-11-20.3 &  24 & 1.07 & 0.80 & R & UH88 & Dundon &      \citet{Ansdell2014} \\
 35 & 2013-11-23.3 &  16 & 1.12 & 0.84 & R & UH88 & Ansdell &      \citet{Ansdell2014} \\
 36 & 2013-12-03.2 &  20 & 1.26 & 1.02 & R & Lowell & Meech, Ansdell &      \citet{Ansdell2014} \\
 37 & 2013-12-11.3 &  36 & 1.37 & 1.18 & R & UH88 & Ansdell &      \citet{Ansdell2014} \\
 38 & 2014-11-27.3 &  89 & 1.82 & 0.85 &   & CS3-PDS & Warner &      \citet{Warner2015} \\
 39 & 2014-11-28.2 &  84 & 1.81 & 0.85 &   & CS3-PDS & Warner &      \citet{Warner2015} \\
 40 & 2014-11-28.4 &  58 & 1.81 & 0.84 &   & CS3-PDS & Warner &      \citet{Warner2015} \\
 41 & 2014-11-29.3 &  82 & 1.80 & 0.84 &   & CS3-PDS & Warner &      \citet{Warner2015} \\
 42 & 2014-11-29.5 &  27 & 1.80 & 0.84 &   & CS3-PDS & Warner &      \citet{Warner2015} \\
 43 & 2014-12-10.1 &  91 & 1.71 & 0.78 & R & C2PU & Devog\`{e}le &        This work      \\
 44 & 2014-12-11.9 &  92 & 1.69 & 0.77 & R & C2PU & Devog\`{e}le &        This work      \\
 45 & 2014-12-14.2 &  52 & 1.67 & 0.77 &   & CS3-PDS & Warner &      \citet{Warner2015} \\
 46 & 2014-12-15.3 &  73 & 1.66 & 0.77 &   & CS3-PDS & Warner &      \citet{Warner2015} \\
 % one 4-point lc was rejected and is not listed here
 47 & 2015-01-13.9 &  54 & 1.32 & 0.83 & R & C2PU & Rivet, Hanu\v s, Delbo' &    This work     \\
 48 & 2015-01-17.9 &  50 & 1.27 & 0.85 & V & C2PU & Devog\`{e}le &        This work     \\
 49 & 2015-02-09.8 &  30 & 0.91 & 0.89 & V & C2PU & Devog\`{e}le &         This work   \\
 50 & 2015-02-10.8 &  41 & 0.89 & 0.89 & V & C2PU & Devog\`{e}le &        This work    \\
 51 & 2015-02-11.8 &  39 & 0.87 & 0.89 & V & C2PU & Devog\`{e}le &        This work   \\
 52 & 2015-08-21.6 &  26 & 2.15 & 2.09 & R & UH88 & Bolin &    This work \\
 53 & 2015-09-08.6 &  22 & 2.24 & 1.90 & R & UH88 & Bolin &     This work \\
 54 & 2015-09-09.6 &  30 & 2.24 & 1.89 & R & UH88 & Bolin &      This work \\
 55 & 2015-10-08.5 &  21 & 2.33 & 1.60 & R & UH88 & Bolin &     This work \\
\hline
\end{tabular}
\end{table*}

%\scriptsize{
\begin{table*}
\caption{\label{tab:IR}Thermal infrared measurements available for Phaethon. For each measurement, the table gives the light-time-corrected epoch in Julian date, wavelength $\lambda$, filter, flux with its error, asteroid distances to the Sun $r$ and Earth $\Delta$, and reference to the source.}
\centering
\begin{tabular}{r rrrr rr r}
\hline 
Epoch (LT corr) & $\lambda$ & Filter & \multicolumn{1}{c} {Flux} & \multicolumn{1}{c} {Flux err} & \multicolumn{1}{c} {$\Delta$} & \multicolumn{1}{c} {$r$} & Reference \\
 \multicolumn{1}{c} {JD} & \multicolumn{1}{c} {[$\mu m$]} &  & \multicolumn{1}{c} {[$Jy$]} & \multicolumn{1}{c} {[$Jy$]} & \multicolumn{1}{c} {au} & \multicolumn{1}{c} {au} &  \\
\hline\hline
2\,445\,618.566846 & 12.0 & I1 & 2.826 & 0.317 & 1.028 & 0.370 &       IRAS \\
2\,445\,618.566846 & 25.0 & I2 & 3.453 & 0.563 & 1.028 & 0.370 &       IRAS \\
2\,445\,618.566846 & 60.0 & I3 & 1.101 & 0.250 & 1.028 & 0.370 &       IRAS \\
2\,445\,618.781563 & 12.0 & I1 & 2.377 & 0.272 & 1.032 & 0.371 &       IRAS \\
2\,445\,618.781563 & 25.0 & I2 & 3.617 & 0.610 & 1.032 & 0.371 &       IRAS \\
2\,445\,618.781563 & 60.0 & I3 & 1.161 & 0.241 & 1.032 & 0.371 &       IRAS \\
2\,445\,618.781563 & 100.0 & I4 & 0.445 & 0.096 & 1.032 & 0.371 &       IRAS \\
2\,445\,618.853136 & 12.0 & I1 & 2.143 & 0.245 & 1.033 & 0.371 &       IRAS \\
2\,445\,618.853136 & 25.0 & I2 & 3.601 & 0.594 & 1.033 & 0.371 &       IRAS \\
2\,445\,618.853136 & 60.0 & I3 & 1.232 & 0.279 & 1.033 & 0.371 &       IRAS \\
2\,445\,618.924708 & 12.0 & I1 & 2.164 & 0.257 & 1.034 & 0.371 &       IRAS \\
2\,445\,618.924708 & 25.0 & I2 & 3.024 & 0.493 & 1.034 & 0.371 &       IRAS \\
2\,445\,618.924708 & 60.0 & I3 & 1.212 & 0.249 & 1.034 & 0.371 &       IRAS \\
2\,445\,618.996292 & 12.0 & I1 & 1.964 & 0.242 & 1.035 & 0.371 &       IRAS \\
2\,445\,618.996292 & 25.0 & I2 & 3.275 & 0.483 & 1.035 & 0.371 &       IRAS \\
2\,445\,618.996292 & 60.0 & I3 & 1.087 & 0.219 & 1.035 & 0.371 &       IRAS \\
2\,445\,619.067876 & 12.0 & I1 & 1.852 & 0.255 & 1.036 & 0.372 &       IRAS \\
2\,445\,619.067876 & 25.0 & I2 & 3.116 & 0.455 & 1.036 & 0.372 &       IRAS \\
2\,445\,619.067876 & 60.0 & I3 & 1.007 & 0.237 & 1.036 & 0.372 &       IRAS \\
\hline
2\,446\,054.797880 & 10.6 & N & 3.853 & 0.231 & 1.131 & 0.246 & \citet{Green1985} \\
2\,446\,054.802080 & 19.2 & Q & 5.380 & 0.323 & 1.131 & 0.246 & \citet{Green1985} \\
2\,446\,054.806880 & 4.7 & M & 0.174 & 0.030 & 1.131 & 0.246 & \citet{Green1985} \\
2\,446\,054.810380 & 8.7 & - & 2.566 & 0.103 & 1.131 & 0.246 & \citet{Green1985} \\
2\,446\,054.812480 & 9.7 & - & 3.175 & 0.190 & 1.131 & 0.246 & \citet{Green1985} \\
2\,446\,054.815280 & 10.3 & - & 3.326 & 0.200 & 1.131 & 0.246 & \citet{Green1985} \\
2\,446\,054.817380 & 11.6 & - & 4.296 & 0.215 & 1.131 & 0.246 & \citet{Green1985} \\
2\,446\,054.820780 & 12.5 & - & 4.436 & 0.266 & 1.131 & 0.246 & \citet{Green1985} \\
2\,446\,054.822880 & 10.6 & N & 3.450 & 0.207 & 1.131 & 0.246 & \citet{Green1985} \\
2\,446\,054.854181 & 10.6 & N & 3.646 & 0.219 & 1.130 & 0.246 & \citet{Green1985} \\
%2\,446\,054.903481 & 3.5 & L & 0.021 & 0.001 & 1.130 & 0.246 & \citet{Green1985} \\
%2\,446\,054.906881 & 3.7 & L' & 0.033 & 0.002 & 1.130 & 0.246 & \citet{Green1985} \\
2\,446\,054.913881 & 4.7 & M & 0.194 & 0.016 & 1.129 & 0.246 & \citet{Green1985} \\
%2\,446\,055.826386 & 3.5 & L & 0.022 & 0.001 & 1.115 & 0.245 & \citet{Green1985} \\
%2\,446\,055.829186 & 3.7 & L' & 0.042 & 0.002 & 1.115 & 0.245 & \citet{Green1985} \\
2\,446\,055.831186 & 4.7 & M & 0.221 & 0.018 & 1.115 & 0.245 & \citet{Green1985} \\
\hline
\end{tabular}
%\tablefoot{
%}
\end{table*}
%\onecolumn
%\scriptsize{
\begin{table*}
\caption{\label{tab:param}Rotation state parameters derived for Phaethon by the light-curve inversion from different photometric datasets. The table gives the ecliptic coordinates $\lambda$ and $\beta$ of all possible pole solutions, the sidereal rotational period $P$, and reference. The uncertainties in our pole solutions are 5 degrees.}
\centering
\begin{tabular}{cccc c c}
\hline
\multicolumn{1}{c} {$\lambda_1$} & \multicolumn{1}{c} {$\beta_1$} & \multicolumn{1}{c} {$\lambda_2$} & \multicolumn{1}{c} {$\beta_2$} & \multicolumn{1}{c} {$P$} & Note \\
\multicolumn{1}{c} {[deg]} & \multicolumn{1}{c} {[deg]} & \multicolumn{1}{c} {[deg]} & \multicolumn{1}{c} {[deg]} & \multicolumn{1}{c} {[hours]} &  \\
\hline\hline
 319 & $-$39 & 84 & $-$39 & 3.603958$\pm$0.000002 & This work \\
 %318$\pm$5 & $-$39$\pm$5 & 84$\pm$5 & $-$39$\pm$5 &  &  &  &  & 3.603958$\pm$0.000002 & Dense data + Catalina \\
 %316$\pm$10 & $-$32$\pm$10 & 82$\pm$10 & $-$31$\pm$10 & 2$\pm$10 & $-$27$\pm$10 & 197$\pm$10 & $-$54$\pm$10 & 3.603958$\pm$0.000002 & Dense data + Lowell \\
  &  & 85$\pm$13 & $-$20$\pm$10 & 3.6032$\pm$0.0008 & \citet{Ansdell2014} \\
 276 & $-$15 & 97 & $-$11 & 3.59060 & \citet{Krugly2002} \\
\hline
\end{tabular}
%*Preferred solution by \citet{Marchis2014}.
%\tablefoot{
%}
\end{table*}
%\onecolumn
\begin{table*}
\caption{\label{tab:TPM}Thermophysical properties of asteroid Phaethon derived by the TPM from different thermal datasets based on two pole solutions. We also include the results based on the varied shape TPM approach (only first pole solution). The table provides the volume-equivalent diameter $D$, thermal inertia $\Gamma$, visual geometric albedo $p_\mathrm{V}$, macroscopic surface roughness, the best-fit $\chi^2_{\mathrm{red}}$, heliocentric distance $r$ of Phaethon during its observation, and the thermal dataset used.}
\centering
\setlength{\extrarowheight}{.5em}
\begin{tabular}{ccccccc}
\hline 
 \multicolumn{1}{c} {$D$} & \multicolumn{1}{c} {$\Gamma$} & \multicolumn{1}{c} {$p_\mathrm{V}$} & \multicolumn{1}{c} {Roughness} & \multicolumn{1}{c} {$\chi^2_{\mathrm{red}}$} & \multicolumn{1}{c} {$r$}  & \multicolumn{1}{c} {Dataset} \\
 \multicolumn{1}{c} {[km]} & \multicolumn{1}{c} {[\tiu]} &  &  &  & \multicolumn{1}{c} {[au]} &  \\
\hline\hline

\multicolumn{7}{c} {Pole 1: $\lambda$=319$^{\circ}$, $\beta$=$-$39$^{\circ}$} \\
6.0$^{+0.5}_{-0.3}$ &   $-$ &  0.09$^{+0.04}_{-0.2}$ & $-$ & 1.0 & 1.0 & IRAS \\
4.6$^{+0.4}_{-0.2}$ &   300$^{+400}_{-100}$ &   0.145$^{+0.008}_{-0.013}$ &              Medium & 1.1 & 1.1 & \citet{Green1985} \\
5.1$^{+0.2}_{-0.2}$ &   600$^{+200}_{-200}$ &   0.122$^{+0.008}_{-0.008}$ &              Medium & 2.8 & 1.1 & Spitzer \\
5.1$^{+0.2}_{-0.2}$ &   600$^{+200}_{-200}$ &   0.122$^{+0.008}_{-0.008}$ &              Medium & 2.9 & $\sim$1.0 & All \\ 
5.1$^{+0.3}_{-0.3}$ &   700$^{+300}_{-300}$ &   0.12$^{+0.01}_{-0.01}$ & Medium & 2.9 & $\sim$1.0 & VS-TPM All \\ 

\multicolumn{7}{c} {Pole 2: $\lambda$=84$^{\circ}$, $\beta$=$-$39$^{\circ}$} \\
5.6$^{+0.7}_{-0.5}$ &   $-$ & 0.10$^{+0.15}_{-0.02}$ &  $-$ & 0.7 & 1.0 & IRAS \\
5.3$^{+0.5}_{-0.4}$ &   4000$^{+2\,000}_{-2\,000}$ &    0.13$^{+0.02}_{-0.02}$ & High & 1.2 & 1.1 & \citet{Green1985} \\
5.2$^{+0.2}_{-0.2}$ &   6500$^{+3\,500}_{-1\,000}$ &  0.14$^{+0.01}_{-0.01}$ & $-$ & 2.5 & 1.1 & Spitzer \\
4.9$^{+0.2}_{-0.2}$ &   7500$^{+2\,500}_{-3\,000}$ &   0.15$^{+0.01}_{-0.01}$ & $-$ & 3.0 & $\sim$1.0 & All \\ 
\hline
\end{tabular}
%\tablefoottext{a}{J\,m$^{-2}$\,s$^{-1/2}$\,K$^{-1}$.}
%\tablefoottext{2}{2.} }
%\footnotetext[2]{Text}

\end{table*}
\bibliography{mybib}
\bibliographystyle{aa}

\end{document}